\documentclass{ws-ijmpa}

\usepackage{mathrsfs}
\newcommand{\beq}{\begin{equation}}
\newcommand{\eeq}{\end{equation}}
\newcommand{\bq}{\begin{equation}}
\newcommand{\eq}{\end{equation}}
\newcommand{\ba}{\begin{array}}
\newcommand{\ea}{\end{array}}
\newcommand{\beqa}{\begin{eqnarray}}
\newcommand{\eeqa}{\end{eqnarray}}

\newcommand{\dif}{\partial}

\def\ep{\epsilon}

\def\zetah{\widehat{\zeta}}
\def\aa{\mathbf{a}}
\def\A{{\cal A}}

\def\D{{\cal D}}

\def\T{{\cal T}}

\def\X{{\mathcal X}}

\def\End{\end{document}}

\def\to{\rightarrow}

\def\dis{\displaystyle}
\def\f{\frac}
\def\ov{\overline}

\def\[{\left[}
\def\]{\right]}
\def\({\left(}
\def\){\right)}

\def\under{\underline}
\def\und{\underline}

\def\Ah{\widetilde{A}}

\def\At{\widetilde{A}}
\def\Bt{\widetilde{B}}
\def\pit{\widetilde{\pi}}
\def\l{{\ell}}

\def\LL{{\mathcal L}}
\def\LLH{\widehat{\mathcal{L}}}

\def\U1EM{U(1)_{\rm em}}

\def\leqq{\leqslant}
\def\geqq{\geqslant}
\def\lan{\langle}
\def\ran{\rangle}
\def\gwt{\widetilde{g}}
\def\fwt{\widetilde{f}}

\def\cb{\cos\beta}

\def\d{\delta}
\def\gt{\widetilde{g}}
\def\[{\left[}
\def\]{\right]}
\def\dis{\displaystyle}

\def\cut{\Lambda}

\def\ot{\otimes}
\def\Es{E^{\star}}
\def\Essm{E^{\star}_{\rm SM}}

\def\ffb{f_2^{~}}
\def\ffc{f_3^{~}}

\def\sh{\widehat{s}}
\def\cb{\overline{c}}
\def\Ch{\widehat{C}}
\def\xh{\widehat{x}}

\def\MT{\widetilde{M}}

\def\Nt{\widetilde{N}}
\def\onet{\widetilde{1}}
\def\zerot{\widetilde{0}}
\def\ft{\widetilde{f}} 
\def\CC{{\mathcal C}}
\def\Sh{\widehat{S}}

\def\Vnab{\underline{V_n^a}}
\def\VVB{\mathbb{V}}
\def\VVBb{\underline{\mathbb{V}}}
\def\Kb{\underline{K}}

\def\KKBb{\underline{\mathbb{K}}}
\def\DDB{\mathbb{D}}
\def\CDb{\underline{\mathcal D}}

\def\Sc{\mathcal{S}}

\def\Kb{\underline{K}}
\def\Omegab{\underline{\Omega}}
\def\Omegahb{\underline{\widehat{\Omega}}}
\def\Fh{\widehat{F}}
\def\FB{\overline{F}}

\def\PIT{\widetilde{\Pi}}

\def\RRB{\mathbb{R}}
\def\RRBT{\widetilde{\mathbb{R}}}
\def\AAB{\mathbb{A}}
\def\WWB{\mathbb{W}}

\def\XXB{\mathbb{X}}
\def\XXBT{\widetilde{\mathbb{X}}}
\def\XT{\widetilde{X}}
\def\XXT{\widetilde{\mathcal X}}
\def\CF{\mathcal F}
\def\CG{\mathcal G}
\def\CD{\mathcal D}
\def\CCB{\mathbb{C}}
\def\CBB{\mathbf{C}}

\def\CCBbar{\overline{\mathbb{C}}}
\def\ch{\widehat{c}}
\def\cbh{\widehat{\overline{c}}}

\def\JJBb{\underline{\mathbb{J}}}
\def\IIB{\mathbb{I}}
\def\IIBbar{\overline{\mathbb{I}}}
\def\ZZB{\mathbb{Z}}

\def\FFB{\mathbb{F}}
\def\FFBu{\overline{\mathbb{F}}}
\def\QQB{\mathbb{Q}}
\def\MMB{\mathbb{M}}
\def\MMW{\mathbb{M}_{W}^{\,2}}
\def\MMZ{\mathbb{M}_{Z}^{\,2}}

\def\MMWT{\widetilde{\mathbb{M}}_{W}^{\,2}}
\def\MMZT{\widetilde{\mathbb{M}}_{Z}^{\,2}}

\def\MMBT{\widetilde{\mathbb{M}}}

\def\Pit{\widetilde{\Pi}}
\def\pit{\widetilde{\pi}}
\def\gh{\widehat{g}}

\def\Ah{\widehat{A}}

\def\AB{\mathbf{A}}

\def\Phip{\Phi_{\alpha}}
\def\Phihp{\widehat{\Phi}_{\alpha}}

\def\pib{\mbox {\boldmath$\pi$}}
\def\Ab{\mathbf A}

\def\aa{\mathbf a}

\def\ka{\kappa}
\def\D{\Delta}
\def\DeT{\widetilde{\Delta}}
\def\ep{\epsilon}
      
\def\Es{E^{\star}}
\def\Du{\mathscr{D}_{\rm U}^{~}}

\def\Duu{\mathscr{D}_{\rm U}}
\newcommand{\dfrac}[2]{\frac{\strut \displaystyle{#1}}
                      {\strut \displaystyle{#2}}}
\newcommand{\df}[2]{\frac{\strut \displaystyle{#1}}
                      {\strut \displaystyle{#2}}}

\begin{document}

\markboth{H.-J. He}%
{Higgsless Deconstruction Without Boundary Condition}

%
\catchline{}{}{}{}{}  
%

\title{\hspace*{-10mm}
       HIGGSLESS DECONSTRUCTION WITHOUT BOUNDARY CONDITION
$\!$\footnote{Presented at DPF-2004: Annual Meeting of
the Division of Particles and Fields, American Physical 
Society, Riverside, California, 
USA, August 26-31, 2004. 
}\hspace*{-10mm}
}

\author{\normalsize {\sc Hong-Jian He}}

\address{\small Department of Physics, 
                University of Texas at Austin, 
                TX 78712, USA\\[1mm]
(\,Email: hjhe@physics.utexas.edu\,) }

\maketitle

\pub{Received: 26 November, 2004}{} 

\begin{abstract}
Deconstruction is a powerful means to explore the rich dynamics 
of gauge theories in four and higher dimensions. 
We demonstrate that gauge symmetry breaking in a compactified
higher dimensional theory can be formulated via deconstructed 
4D moose theory with {\it spontaneous symmetry breaking} and 
{\it without boundary condition.}  The proper higher-D boundary 
conditions are automatically induced in the continuum limit 
rather than being imposed. We identify and analyze the moose  
theories which exhibit {\it delayed unitarity violation} 
(effective unitarity) as a {\it collective effect} of many 
gauge groups, without resorting to any known 5D geometry. 
Relevant phenomenological constraints are also addressed.
\hfill [hep-ph/0412113]

\keywords{Deconstruction, Unitarity, 
          Higgsless Electroweak Symmetry Breaking}

\end{abstract}

\section{Advantageous Deconstruction}

\renewcommand{\thefootnote}{\arabic{footnote}}

Deconstruction\cite{DC} 
is a powerful means to explore the rich
dynamics of gauge theories in four and higher dimensions. 
A compactified higher dimensional theory may be deconstructed
into a proper moose representation\cite{moose} or the
equivalent transverse Wilson lattice\cite{Tlattice}. 
An essential advantage of the deconstruction  
is that it allows us to formulate the often involved 
higher dimensional gauge symmetry breaking (without/with 
gauge group rank reduction) in terms
of the conventional 4D gauged nonlinear sigma model \`{a} la 
CCWZ\cite{CCWZ}, 
where no extra boundary condition (BC) is required {\it a priori.} 
Furthermore, deconstruction allows us to identify and analyze 
the general 4D moose theories with arbitrary inputs of gauge
couplings $g_j^{~}$ and Goldstone decay constants $f_k^{~}$ 
or with only a few Kaluza-Klein (KK)\cite{KK} modes. 
Such theories need not resemble any known higher-D geometry,
but as we will show,  
they can still exhibit {\it delayed unitarity violation}
(effective unitarity) as a {\it collective effect} of many 
participating gauge groups,  which was originally revealed
for the deconstruction of specified 5D geometries\cite{CH,SUSY03}.
The effective unitarity in the compactified or deconstructed
5D Yang-Mills theories is ensured by {\it the presence of spin-1 
vector-bosons (gauge KK modes)}\,\cite{CDH,CH,SUSY03} 
rather than the conventional Higgs scalar\cite{Higgs,SM-UB}, 
which is the key for seeking realistic Higgsless models 
of electroweak symmetry breaking (EWSB) 
in the recent literature\cite{HLxx,HLmoose}.\,
These new vector bosons also provide discovery signatures    
at the CERN LHC.

\section{Eaten Goldstone Bosons 
         in General Higgsless Moose Theory}

Consider a most general linear moose theory, 
consisting of the replicated gauge group 
\,$G_A^{N+1}\otimes G_B^{M+1}\otimes G_C^{L+1}\otimes\cdots$\, 
which is {\it spontaneously broken} to the diagonal
subgroup $\,G_D\,$ by the link fields $\,U_j\,$  
transforming as 
bi-fundamentals under the two adjacent gauge groups. 
Without losing generality,  we will set\cite{HLmoose}
\,$G_A=SU(2)$,\, \,$G_B=U(1)$\, and \,$G_C = \cdots =0$\,
for convenience of analyzing the EWSB,
so the residual gauge symmetry is 
$\,G_D=U(1)_{\rm em}\,$\, with a massless photon,   
as depicted in Fig.\,1\,\cite{HLmoose}.

\begin{figure}[h]
\label{fig:Fig1}
\begin{center}
\vspace*{-5mm}
\centerline{\psfig{file=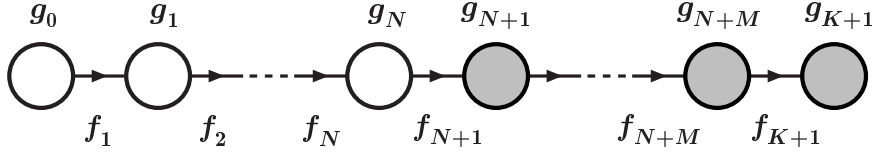,width=13cm}}
\setlength{\unitlength}{1mm}
\begin{picture}(0,0)
\end{picture}
\vspace*{-5mm}
\caption{A generic Higgsless deconstruction of linear moose theory
with arbitrary product gauge group 
\,$G_A^{N+1}\otimes G_B^{M+1}$\,,\, where the
most general inputs of gauge couplings and Goldstone decay constants
$\{g_j^{~},\,f_k^{~}\}$ $(j=0,1,\cdots,K+1;~k=1,2,\cdots,K+1; 
~K\equiv N+M)$ are allowed. 
The open circles represent the group $G_A$
and shaded circles the group $G_B$.}
\end{center}
\vspace*{-4mm}
\end{figure}

The Lagrangian for this linear moose theory is 
\begin{equation}
\label{eq:L}
\dis
  {\cal L} \,~=~\,
  \sum_{j=0}^{K+1} -\dfrac{1}{\,2\,} 
                   \mbox{Tr}\(F_{j\mu\nu} F_j^{\mu\nu}\)
~+\,
  \sum_{j=1}^{K+1} \df{\,f_j^2\,}{4} \mbox{Tr}\[
                   (D_\mu U_j)^\dagger (D^\mu U_j) \],
\end{equation}
with $~
\dis
  D^\mu U_j = \partial^\mu U_j - i g_{j-1}^{~}\Ab_{j-1}^\mu U_j 
                               + i g_j^{~}U_j \Ab_{j}^\mu \,,
~$ 
and
$~U_j = \exp\[i2\pib_j^{~}/f_j^{~}\] \,$,\,
where  
\,$\Ab_j^\mu\equiv A_j^{a\mu}T^a \in SU(2)_j$\, 
for $j=0,1,\cdots, N$,  
and 
\,$\Ab^\mu_j\equiv A^\mu_jT^3\in U(1)_j$ $(j=N+1,\cdots,K+1)$.\,  
Also, 
\,$\pib_j \equiv \pi^a_jT^a$  ($1\leqq j \leqq N+1$)\,
and
\,$\pib_j \equiv \pi^3_jT^3$  ($N+2 \leqq j \leqq K+1$).\,
The $(N+1)\times (N+1)$ mass matrix of charged gauge bosons is
given by\cite{HLmoose}
{\small
\begin{equation}
\label{eq:M_CC}
\MMB_{\,W}^{\,2} = \df{1}{4}
\left(
\begin{array}{c|c|c|c|c}
&&&&
\\[-2mm]
g_0^2 f_1^2    & -g_0^{~} g_1^{~} f_1^2     &      &  
\\[1.5mm]
\hline
&&&&
\\[-2mm]
-g_0^{~} g_1^{~} f_1^2  &  g_1^2(f_1^2\!+\!f_2^2) 
& -g_1^{~} g_2^{~} f_2^2         &  
\\[1.5mm]
\hline
&&&&
\\[-2mm]
& -g_1^{~} g_2^{~} f_2^2     &  g_2^2 (f_2^2 \!+\! f_3^2) & 
                 - g_2^{~} g_3^{~} f_3^2                                
\\[1.5mm]
\hline
&&&&
\\[-4mm]
 & & \ddots & \ddots & \ddots    
\\[0.5mm]
\hline
&&&&
\\[-2mm]
 & & & -g_{N-1}^{~} g_{N}^{~} f_N^2         &
          g_N^2(f_N^2\!+\!f_{N+1}^2)       
\\[1.5mm]
\end{array}
\right),
\end{equation}
}
\noindent\hspace*{-2mm}
which may be denoted as $\,\MMW =\MMB_{[0,N+1)}^{\,2}\,$
by using the notation of open/closed intervals\cite{HLmoose}.
Similarly the $(K+2)\times (K+2)$ mass matrix of neutral
gauge bosons is 
$\,\MMZ =\MMB_{[0,K+1]}^{\,2}\,$  and can be obtained
from the matrix $\,\MMB_{[0,N+1)}^{\,2}\,$ 
with simple replacements,
$\,N\to K+1\,$ and \,$f_{K+2}\to 0$\,.\,
The eigenvalues of $\MMW$ are denoted by
$M_W^2$ for light $W^\pm_0\equiv W^\pm$ boson and 
$M_{Ww}^2$ ($w=1,2,\cdots,N$) for heavy ``KK'' states $W^{\pm}_w$, 
while $\MMZ$ has eigenvalues $0$ ($M_Z^2$) for photon $A_\gamma^0$
(light $Z^0_0\equiv Z^0$ boson) and $M_{Zz}^2$ ($z=1,2,\cdots,K$) 
for heavy ``KK'' states $Z^0_{z}$. 
With the diagonal matrices 
\,$\CF_{n+1}
   ={\rm diag}\(f_1^{~},f_2^{~},\cdots,f_{n+1}^{~}\)/2$\, and
\,$\CG_{n+1}
   ={\rm diag}\(g_0^{~},g_1^{~},\cdots,g_n^{~}\)$\,,\,
it is convenient to define the notations,
$\,\QQB_{\,W} =\CF_{N+1}\CD_W\CG_{N+1}\,$ and
$\,\QQB_Z =\CF_{K+1}\CD_Z\CG_{K+2}\,$, where the 
$(N+1)\times (N+1)$ and $(K+1)\times (K+2)$ 
non-diagonal matrices $\CD_W$ and $\CD_Z$ are
\beq
\label{eq:D}
\CD_W = \dis 
\left(
\begin{array}{c|r|c|c|r}
&&&&
\\[-3.5mm]
\,1\,   &  -1     &    & & 
\\[0mm]
\hline
&&&&
\\[-4.5mm]
  & 1  & -1  && 
\\[0mm]
\hline
&&&&
\\[-5mm]
 & & \ddots & \ddots   &   
\\[0mm]
\hline
&&&&
\\[-4.5mm]
 & &  & 1         &   -1 
\\[0mm]
\hline
&&&&
\\[-4mm]
 & &  & &   1         
\\[-1mm]
\end{array}
\right),  
~~~~~~~~~~
\CD_Z = \dis 
\left(
\begin{array}{c|r|c|c|r|c}
&&&&
\\[-3.5mm]
\,1\,   &  -1     &    & & & 
\\[0mm]
\hline
&&&&&
\\[-4.5mm]
  & 1  & -1  &&& 
\\[0mm]
\hline
&&&&&
\\[-5mm]
 & & \ddots & \ddots   &&   
\\[0mm]
\hline
&&&&&
\\[-4.5mm]
 & &  & 1         &   -1 & 
\\[0mm]
\hline
&&&&
\\[-4mm]
 & &  & &   1    & -1         
\\[-1mm]
\end{array}
\right).
\eeq
So the mass matrices ($\MMW,\,\MMZ$) and 
their duals ($\MMWT,\,\MMZT$) 
can be written as\cite{HLmoose,dualm} 
\beq
\ba{llll}
\MMW = \QQB_W^T\QQB_W^{~}\,,~~ &
\MMZ = \QQB_Z^T\QQB_Z^{~}\,;~~ &
\MMWT = \QQB_W^{~}\QQB_W^T\,,~~ &
\MMZT = \QQB_Z^{~}\QQB_Z^T\,;
\ea
\eeq
where the dual mass matrices \,$\MMWT$ and $\MMZT$\, share
the same nonzero eigenvalues as 
\,$\MMW$\, and\, $\MMZ$\,,\, respectively. 
This is because $\MMB_a^2$ and $\MMBT_a^2$
($a=W,Z$) are real symmetric matrices 
and $\QQB_a$ can be diagonalized by the
bi-orthogonal rotations, 
\beq
\label{eq:Q-diag}
\RRBT_a^T\QQB_a\RRB_a = \QQB_a^{\rm diag} \,,
\eeq
which leads to
\beq
\label{eq:M2-diag}
\ba{ll}
\RRB_a^T\MMB_a^2\RRB_a 
= (\QQB_a^{\rm diag})^T(\QQB_a^{\rm diag})
\equiv (\MMB_a^2)^{\rm diag} \,,
\\[2mm]
\RRBT_a^T\MMBT_a^2\RRBT_a 
= (\QQB_a^{\rm diag})(\QQB_a^{\rm diag})^T
\equiv (\MMBT_a^2)^{\rm diag} \,.
\ea
\eeq
The $(N+1)\times(N+1)$ matrices
$\MMB_W^2$ and $\MMBT_W^2$ must have identical eigenvalues
(which are all nonzero) because of 
$\,(\MMW )^{\rm diag} = 
   (\QQB_W^{\rm diag})^T(\QQB_W^{\rm diag})=
   (\QQB_W^{\rm diag})(\QQB_W^{\rm diag})^T
  =(\MMWT)^{\rm diag}\,$.\,
On the other hand, the $(K+2)\times(K+2)$ matrix
$\,(\QQB_Z^{\rm diag})^T(\QQB_Z^{\rm diag})= 
   (\MMZ)^{\rm diag}\,$
has one zero-eigenvalue corresponding to the photon mass, 
and it is thus clear that the $(K+1)\times(K+1)$ matrix
$\,(\QQB_Z^{\rm diag})(\QQB_Z^{\rm diag})^T
  =(\MMZT)^{\rm diag}\,$
has $K+1$ massive eigenvalues identical to those in
$(\MMZ)^{\rm diag}\,$.\,

Expanding the Lagrangian (\ref{eq:L}) gives 
the gauge-Goldstone boson mixing term,\cite{HLmoose}
\beq
\label{eq:A-GB}
\ba{lcl}
{\cal L}_{\rm GB}^{\rm mix} &~=~&
\[-{{\mathbb A}^{+}_{\mu}}^T\QQB_{\,W}^{\,T}
    \partial^\mu\Pi^- + {\rm h.c.} \]
  \,-\,{{\mathbb A}^{3}_{\mu}}^T\QQB_{\,Z}^{\,T}
    \partial^\mu\Pi^0 
\\[2mm]
&~=~&
\[-{\WWB^{+}_{\mu}}^T\MMB_{W}^{\rm \,diag}
    \partial^\mu\Pit^- + {\rm h.c.} \]
 \,-\,{\ZZB^{0}_{\mu}}^T\MMBT_{Z}^{\rm \,diag}
    \partial^\mu\Pit^0 \,,
\ea
\eeq
with
\beq
\ba{ll}
\AAB^{\pm\mu}\,=\,
(A_0^{\pm\mu},\,A_1^{\pm\mu},\,\cdots,\,A_N^{\pm\mu}\,)^T,
~~&~~
\AAB^{3\mu}\,=\,
(A_0^{3\mu},\,A_1^{3\mu},\,\cdots,\,A_{K+1}^{3\mu})^T;
\\[2mm]
\WWB^{\pm\mu}\,=\,
(W_0^{\pm\mu},\,W^{\pm\mu}_1,\,\cdots,\,W^{\pm\mu}_N\,)^T, 
~~&~~
\ZZB^{0\mu}\,=\,
(Z_0^{0\mu},\,Z^{0\mu}_1,\,\cdots,\,Z^{0\mu}_{K})^T;
\\[2mm]
\Pi^{\pm}\,=\,
(\pi_1^{\pm},\,\pi^{\pm}_2,\,\cdots,\,\pi^{\pm}_{N+1}\,)^T,
~~&~~
\Pi^{0}\,=\,
(\pi_1^{0},\,\pi^{0}_2,\,\cdots,\,\pi^{0}_{K+1})^T;
\\[2mm]
\Pit^{\pm}\,=\,
(\pit_0^{\pm},\,\pit^{\pm}_1,\,\cdots,\,\pit^{\pm}_N\,)^T,
~~&~~
\Pit^{0}\,=\,
(\pit_0^{0},\,\pit^{0}_1,\,\cdots,\,\pit^{0}_K)^T;
\ea
\eeq
where $\,(\WWB^{\pm\mu},\,\ZZB^{0\mu})\,$
are mass-eigenbasis fields,  while
$\,(\Pit^\pm,\,\Pit^0)\,$ are ``eaten'' Goldstone fields,
which are connected to the site Goldstone states
$\,(\Pi^\pm,\,\Pi^0)\,$ via
\beq
\label{eq:PI-rotation}
\Pit^\pm \,=\, \RRBT_W^T\,\Pi^\pm \,, ~~~~~~
\Pit^0   \,=\, \RRBT_Z^T\,\Pi^0 \,.
\eeq
Hence, the ``eaten'' Goldstones are exactly aligned with the
``gauge boson'' mass-eigenstates of the dual moose.
The gauge-Goldstone mixing (\ref{eq:A-GB}) can be
removed by the familiar 
$R_\xi$ gauge-fixing term, 
\beq
\label{eq:GF}
\ba{l}
{\cal L}_{\rm gf} \,=\, \dis
\sum_{n=0}^N
- \f{1}{\,\xi_W^{~}}F_n^+ F_n^- \,+\,
\sum_{n=0}^{K+1} -\f{1}{\,2\xi_Z^{~}\,}(F_n^0)^2 
\,,
\\[6mm]
\dis
F_n^a \,=\, 
\dif_\mu V_n^{a\mu}  + 
\xi_a M_{an}\pit_n^a  \,,
\ea
\eeq
where 
$\,V_n^{a\mu}\in (W_n^{\pm\mu},\,Z_n^{0\mu}),~
   M_{an}\in (M_{Wn},M_{Zn})$\, and
$\,\xi_a^{~}\in (\xi_W^{~},\,\xi_Z^{~})\,$.\,
Also, $\,Z^{0\mu}_{K+1}\equiv A_{\gamma}^{0\mu}\,$,\,
$\,M_{Z(K+1)}\equiv M_{\gamma}=0\,$ and 
$\,\pit^0_{K+1} \equiv 0\,$.\,
From Eq.\,(\ref{eq:GF}), the ``eaten''  Goldstone boson mass is 
given by  $~M_{\pit^a_n}^2=\xi_a^{~}M_{an}^2\,$.

\vspace*{3mm}
\section{4D Higgs Mechanism and Geometrization 
         in 5D Continuum}

Corresponding to the gauge-fixing (\ref{eq:GF}), we
derive Faddeev-Popov ghost term, 
\beq
\label{eq:FP}
{\cal L}_{\rm FP} \,=\, \dis
\sum_{n=0}^N 
\[\ov{c}_n^+\sh F_n^- + \ov{c}_n^-\sh F_n^+\]
+ \sum_{n=0}^{K+1} 
\ov{c}_n^Z\sh F_n^0 \,, 
\eeq
which ensures BRST\cite{BRST} invariance
at quantum level.
Extending the notations of [\refcite{ET}], we write the BRST 
transformations for the moose theory,
\beq
\label{eq:BRST}
\ba{ll}
\dis
\sh c_n^a      = \dis -\f{1}{2}g_n^{~}C^{abc}c_n^bc_n^c\,,~~~  
&\dis
\sh\,\cb_n^a   = -\xi_a^{-1}F_n^a \,, 
\\[2mm]
\dis
\sh V_n^{a\mu} = D^{a\mu}_{nb}(V)c_n^b \,,~~~
&\dis
\sh \pi_n^a    = D^{a\pi}_{nb}(\pi)c_n^b \,,
\\[2mm]
\dis
\sh \VVB^{a\mu} = \RRB_a^T\DDB^{a\mu}_{b}\CCB_n^b \,,~~~
&\dis 
\sh \PIT^{a} = \RRBT_a^T\DDB^{a\pi}_{b}\CCB_n^b \,,
\ea
\eeq
where 
$\,\VVB^{a\mu}=(V_0^{a\mu},\,\cdots,\,V_{P}^{a\mu})^T\,$,\,
$\,\PIT^{a}=(\pit_0^{a},\,\cdots,\,\pit_{P'}^{a})^T\,$,\,
$\,\CCB^{a}=(c_0^{a},\,\cdots,\,c_P^{a})^T\,$,\,
and $P,P'=N ~(P=K+1,\,P'=K)$ for charged (neutral) fields.
In (\ref{eq:BRST}),
$\,D^{a\mu}_{nb}(V)\,$ and $\,D^{a\pi}_{nb}(\pi)\,$
are given by the gauge transformations of  
\,$V_n^{a\mu}$\, and $\,\pi_n^a$,\, respectively; also 
we define the matrix
$\,\DDB^{i}_b = {\rm diag}(D^i_{0b},\cdots,D^i_{Pb})^T\,$.\,
Following [\refcite{CH,SUSY03}], 
we directly extend 
Appendix-A of the fourth paper in [\refcite{ET}] to
derive a 4D Slavnov-Taylor (ST) identity for the moose
theory (\ref{eq:L}) with replicated gauge group, 
\beq
\label{eq:4D-ETI1}
\lan 0|T 
F^{a_1^{~}n_1^{~}}(x_1^{~})
F^{a_2^{~}n_2^{~}}(x_2^{~})
\cdots
F^{a_\ell^{~}n_\ell^{~}}(x_\ell^{~})\Phip |0\ran \,=\, 0\,,
\eeq 
which is due to the 
{\it manifest gauge invariance} of the moose
action \,$\int dx^4\LL$\, from (\ref{eq:L}) or
the equivalent BRST invariance of the effective action
\,$\int dx^4 \[\LL+\LL_{\rm gf}+\LL_{\rm FP}\]$\,.\,
In (\ref{eq:4D-ETI1}) 
$\,\Phip$\, denotes other possible amputated physical fields. 
We stress that (\ref{eq:4D-ETI1}) generally holds
for {\it any} possible form of the gauge-fixing function 
$\,F^{an}$.\,  
Compared to the SM case\cite{ET}, the main complication
below comes from amputating the external fields $F^a_n$
in (\ref{eq:4D-ETI1}), due to the rotations in
(\ref{eq:Q-diag})-(\ref{eq:M2-diag}).
We find it convenient to introduce a matrix notation for
all relevant gauge-fixing functions,
\beq
\label{eq:F-matrix}
\FFB^a = {\KKBb^a}^T\VVBb^a \,,~~~~~
\VVBb^a \equiv
\(\ba{c}\VVB^{a\mu}  \\[1mm] \Pit^a \ea
\),~~~~~
\KKBb^a \equiv
\(\ba{c} \und{I}_a\dif^\mu \\[1mm] 
         \xi_a \und{M}_a \ea
\),
\eeq
where \,$\und{I}_a={\rm diag}(1,\cdots,1)$\, and
$\,\und{M}_a ={\rm diag}(M_{a0},\cdots,M_{aP})\,$
are $(1+P)\times (1+P)$ diagonal matrices with 
$\,P=N\,(P=K+1)\,$ for charged (neutral) sector. 
In (\ref{eq:F-matrix}) we have extended
the column $\Pit^a$ to \,$1\!+\!P$\, dimensional with 
$\,\pit^0_{K+1}\equiv 0\,$.\,
Introducing the external source term
$\,\int d^4x \[{\JJBb^a}^T\VVBb^a
               +\IIBbar^a\CCB^a
               +\CCBbar^a\IIB^a\]\,$,\,
we derive a generating equation for connected Green 
functions,
\beq
{\JJBb^a}^T\lan 0|T\sh\,\VVBb^a(x)|0\ran 
+\IIBbar^a(x)\lan 0|T\sh\CCB^a(x)|0\ran
-\lan 0|T\sh\CCBbar^a(x)|0\ran\IIB^a(x) ~=~ 0\,,
\eeq
from which we deduce an ST identity for the matrix 
propagator of \,$\VVBb^a$,
\beq
\label{eq:ST-D}
{\KKBb^a}^T\CDb^{ab}(p) = -[\Omegab^{ab}](p)^T   \,,
\eeq
with
\beq
\label{eq:ST-Dx}
\ba{l}
\CDb^{ab}(p) =\lan0|T{\VVBb^a}{\VVBb^b}^T|0\ran(p) \,,~~~
{\mathcal S}(p)\delta^{ab} 
=\lan0|T\CCB^{\,a}\CCBbar^b|0\ran(p) \,,
\\[2mm]
\Omegab^{ab}(p) \equiv \Omegahb^{ab}\!(p)\,
                       {\mathcal S}(p)\,,
\\[3mm]
\Omegahb^{ab}(p) \equiv
\lan 0|T\sh\VVBb^b|\CCBbar^a\ran (p)
\equiv \!
\(\ba{c} \lan 0|T\sh\VVB^{b\mu}|\CCBbar^a\ran
\\[1mm]  \lan 0|T\sh\,\Pit^b|\CCBbar^a\ran 
\ea
\)_{\!\!(p)}  \!\!\! \equiv\!
\(\!\ba{c} 
-ip^\mu\delta^{ab}\RRB_a^T [\mathbf{1}\!+\!\Delta_V^a(p^2)]
\\[1mm]  
\delta^{ab}\RRBT_a^T\QQB_a [\mathbf{1}\!+\!\Delta_\pi^a(p^2)] 
\ea
\)\!.
\ea 
\eeq
Using (\ref{eq:ST-D}) and collecting $F^{an}$ in the matrix 
form $\FFB^a$ for each external line in (\ref{eq:4D-ETI1}), 
we make an amputation for (\ref{eq:4D-ETI1}),
$\,0=G[\FFB^a(p),\cdots] 
    =-[\Omegab^{ab}]^T\T [\VVBb^a(p),\cdots],
\,$
which leads to 
\beq
\ba{cl}
0  & =~
\T [\,p_\mu\!\VVB^{a\mu}(p)-\MMBT_a^{\rm diag}
     {\mathbf C}^a\Pit^a(p),\,
     \cdots\cdots \,]  \,,
\\[3mm]
\CBB^a &\equiv~ -i\,
(\QQB_a^T\RRBT_a)^{-1}
\(\mathbf{1}\!+\!{\Delta_V^a}(p^2)^T\)^{-1} \!
\(\mathbf{1}\!+\!{\Delta_\pi^a}(p^2)^T\) 
(\QQB_a^T\RRBT_a)
\\[3mm]
& =~ -i\[\,\mathbf{1}+O({\rm loop})\,\]
\,.
\ea
\eeq
Repeating this amputation for all external lines in
(\ref{eq:4D-ETI1}), we arrive at a matrix identity
for $S$-matrix elements, 
\beq
\label{eq:4D-ETI2}
 \T\!\[\,\FFBu^{a_1^{~}}(p_1^{~}),\,
         \FFBu^{a_2^{~}}(p_2^{~}),\, \cdots, \,
         \FFBu^{a_{\ell}}(p_{\ell}^{~}),\,\Phip\] 
     \,=~ 0\,,
\eeq
where 
$\,\FFBu^{a}(p)
\equiv p^\mu \VVB_\mu^{a}-\MMBT_a^{\rm diag}\CBB^{a}\Pit^a
=\MMBT_a^{\rm diag}(\VVB_S^{a}-\CBB^{a}\Pit^a)\,$
with $\,\VVB_S^{a}=\ep^\mu_S\VVB_\mu^{a}\,$ 
($\ep_S^\mu \equiv p^\mu/\MMBT_a^{\rm diag}$),\, 
$\VVB^{a\mu} =\WWB^{\pm\mu} ,\,\ZZB^{0\mu}$,\,
and 
\,$\CBB^a = -i[1+O({\rm loop})]$.\,
The identity (\ref{eq:4D-ETI2}) states 
that in the $S$-matrix element
the unphysical eaten ``KK'' Goldstones 
$\Pit^{a}$
and the unphysical ``KK'' scalar gauge-components 
$\VVB_S^{a}$ are {\it confined,} so they together
have no net contribution to any physical process ---
a quantitative formulation of the 4D Higgs mechanism at
the $S$-matrix level, which holds even without a physical 
Higgs boson such as in the present moose theory (gauged 
nonlinear sigma model) with replicated gauge groups.  
Under high energy expansion, 
Eq.\,(\ref{eq:4D-ETI2}) results in 
a {\it generalized form} of
the equivalence theorem (ET)\cite{ET,CH}, which connects the
high energy longitudinal gauge boson scattering amplitude
to that of the ``eaten'' Goldstone bosons, 
\beq
\label{eq:ETm}
\dis
\T\!\[V_L^{a_1^{~}n_1^{~}},
      V_L^{a_2^{~}n_2^{~}},\,
\cdots\,\] =\, C_{\rm mod}^{n_1m_1,n_2m_2\cdots}\,
\T\!\[\pit^{a_1^{~}m_1^{~}},
      \pit^{a_2^{~}m_2^{~}},\,
      \cdots\,\] 
      \,+\, O\!\(\!\f{M_{an}}{E_n}\!\)\!,
\eeq
where $\,V_n^a=W^\pm_{n},\,Z^0_{n}\,$
($n=0,1,2,\cdots$),\, and sums over repeated indices
$(m_1,m_2,\cdots)$ are implied. 
The exact expression of the $\,O(M_{an}/E_n)\,$ term is 
obtained from directly expanding the ET identity 
(\ref{eq:4D-ETI2}),
in the {\it same} way as in [\refcite{ET}].
The radiative modification factor is
\beq
\label{eq:Cmod}
C_{\rm mod}^{n_1m_1,n_2m_2\cdots}
   \equiv\, \CBB^{a_1}_{n_1m_1}\CBB^{a_2}_{n_2m_2}\cdots
   \,=\,(-i)^{\ell} 
        \[(\delta_{n_1m_1}\delta_{n_2m_2}\cdots)+O({\rm loop})\],
\eeq
which extends
Ref.\,[\refcite{ET}] to the case of replicated gauge group.\,

Analyzing both sides of (\ref{eq:ETm}),  
we observed\,\cite{CH} that the non-canceled leading
$E^2$-terms in the usual nonlinear gauged sigma model
are now suppressed by
large $N$ due to the {\it collective effect} of 
many ``KK'' modes in the deconstruction theory.

To take the 5D continuum limit we redefine the 
lattice link field (Wilson line), 
\beq
\label{eq:U-lattice}
U_j(x)=U(x,x^5_j)
 =\exp\!\[i\!\int_{x^5_{j-1}}^{x^5_j}\!\!\!\!\! dx^5 
  g_j^{~}\AB_5(x,x^5)\]\!
 =\exp\!\[i\,\mathbf{a} g_j^{~}\AB_5(x,x^5_j)\]
\eeq
where $\,\mathbf{a}=\Delta x^5_j\,$ is lattice spacing and
\,$A_5^a(x,x^5_j)$\, is related to the site-Goldstone boson 
$\pi^a_j(x)$ in (\ref{eq:L}),
$\,A_5^a(x,x^5_j)
  =\pi^a_j(x)(2/g_j^{~}f_{j\,}\mathbf{a})\,$.\,
So the site-Goldstone field $\pi^{a}_j$  
will geometrize as $\Ah^a_5$ component of the 5D gauge field 
$\,\Ah^a_J = (\Ah^a_\mu,\,\Ah^a_5)\,$,\, 
where $\,J,I,\ldots\in (\mu,\,5)\,$ 
denote the 5D Lorentz indices.
Following [\refcite{DeRS}] and ignoring the $U(1)^{M+1}$
part of Fig.\,1 for simplicity,  
we define the position-dependent couplings 
$\,g_j^{~} = g_5^{~}\kappa_j^{~}/\sqrt{\mathbf a}\,$
and decay constants 
$\,f_j^{~}=h_j^{~}f\,$, where $g_5^{~}$ and $f$ are pure
constants of mass-dimension $-\f{1}{2}$ and $1$, respectively.
Thus we can derive, 
$\,D^{\mu}U_j 
= -i\,\mathbf{a}g_j^{~}F^{a\mu 5}_j
+O(\mathbf{a}^2) 
= -i\,\mathbf{a}g_5^{~}\Fh^{a\mu 5}(x,x^5_j)
+O(\mathbf{a}^2)\,$, where the identification 
$\, A^{a\mu,5}_j(x) \to 
    (\sqrt{\mathbf{a}}/\kappa_j^{~})
    \Ah^{a J}(x,x^5_j)
 = (g_5^{~}/g_j^{~})\Ah^{a J}(x,x^5_j)
$\, 
is made under $\,\mathbf{a}=L/N\to 0\,$.\,
As such, we see that the Goldstone Lagrangian in the second
term of (\ref{eq:L}) just gives the continuum $\Ah^a_5$-Lagrangian,
$\,\int_0^L dx^5 
   \[-\f{1}{2}h^2(x^5)\Fh^{a\mu 5}\Fh^a_{\mu 5}\]\,$,\,
after imposing a normalization condition 
$\,\mathbf{a}=(2/g_5^{~}f)^2\,$.\,
With these, we reproduce a 
bulk $SU(2)_{\rm 5D}$ gauge theory from (\ref{eq:L})
in the continuum limit,
\beq
\label{eq:SU2-5D-L}
S_5^{~} ~=~ \dis\int\! d^5\xh\, \sqrt{-\gh~}~\gh^{IK}\gh^{JL}\,
\f{-1}{\,4\kappa^2(x^5)\,}\,\Fh^a_{IJ}\Fh^a_{KL} \,,
\eeq
defined on a general 5D background,
\beqa
ds^2 &~=~& [\kappa(x^5)h(x^5)]^2\eta_{\mu\nu}^{~}
         dx^\mu dx^\nu -\, dx^5 dx^5 \,,
\eeqa
where the metric 
$\,\gh_{IK}^{~} = 
 {\rm diag}\(\eta_{\mu\nu}^{~}(\kappa h)^2,\,-1\)\,$.\,
For the simplest case $\,\kappa = h =1\,$, it reduces
to flat 5D geometry, while for the case
$\,\kappa =1\,$ and $\,h=\exp [-k|x^5|]\,$ 
it reduces to the familiar warped RS1\cite{RS}.
Various extended continuum models from (\ref{eq:SU2-5D-L}) 
can be derived from the moose theory (\ref{eq:L}) 
by including the $U(1)^{M+1}$ groups and/or proper
folding(s) of the moose chain in Fig.\,1. 
These together with the {\it induced} 5D BCs
and possible brane kinetic terms
will be analyzed in Sec.\,4.
%
%

Analyzing (\ref{eq:SU2-5D-L}) we construct
an $R_\xi$ gauge-fixing term to remove the 
\,$\Ah^{a\mu}-\Ah^{a5}$\, mixing,
\beq
\dis
\LLH_{\rm gf} \,=\, -\f{1}{\,2\xi\,}(\Fh^a)^2\,,
~~~~~\Fh^a =  \kappa^{-1}\dif_\mu\Ah^{a\mu}
            +\xi\kappa\,\dif_5(h^2\Ah^{a5}) \,,
\eeq
The corresponding ghost term is given by
\beq
\label{eq:FP-5D}
\LLH_{\rm FP} ~=~ \dis \cbh^a\sh\Fh^a \,,
\eeq
and the 5D BRST transformations are
\beq
\label{eq:BRST-5D}
\sh\Ah^{aJ}   = D^{aJ}_b(\Ah)\ch^b ,~~~~ 
\sh \ch^a     = \dis-\f{1}{2}g_5^{~}C^{abc}\ch^b\ch^c ,~~~~
\dis
\sh\,\cbh^a   = -\xi^{-1}\Fh^a .
\eeq
Using 5D gauge invariance of the action $\,S_5\,$, we
derive an ST identity\cite{SUSY03} in parallel
to the 4D result (\ref{eq:4D-ETI1}),
\beq
\label{eq:5D-ETI1}
\lan 0|T 
\Fh^{a_1^{~}}(\xh_1^{~})
\Fh^{a_2^{~}}(\xh_2^{~})
\cdots
\Fh^{a_\ell}(\xh_{\ell})\Phihp |0\ran \,=\, 0\,.
\eeq 
The 5D equation of motion (EOM) for free field 
$\,\Ah^{a}_{\mu 5}\,$ is
\beq
\label{eq:5D-EOM}
\ba{l}
\(\dif_{\mu}^2 -\kappa^2\dif_5h^2\dif_5\)\Ah^a_{\nu} -
\(1-\xi^{-1}\)\dif_{\nu}\dif_{\mu}\Ah^{a\mu} \,=\, 0 \,,
\\[2mm]
\dif_5(\kappa^2\dif_5 h^2\Ah^a_5) 
 -\xi^{-1}\dif_{\mu}^2\Ah^a_5 \,=\, 0 \,,
\ea
\eeq
where we have made integration by part in 
the action $S_5+S_{\rm gf}+S_{\rm FP}$ and verified that all
surface terms vanish under the induced consistent BCs
in Sec.\,4.
Then, we define the KK expansions,
\beq
\label{eq:KK-expansion}
\Ah^{a\mu}(\xh) 
=\dis\f{1}{\sqrt{L\,}\,}
 \sum_{n=0}^{\infty} V^{a\mu}_n(x)\X_n(x^5)\,,
~~~\,
\Ah^{a5}(\xh) 
=\dis\f{1}{\sqrt{L\,}\,}
 \sum_{n=0}^{\infty} V^{a5}_n(x)\XXT_n(x^5)\,.
\eeq
As in [\refcite{randall}] we find it convenient to
work in the momentum space for 4D KK fields $V^{a\mu,5}_n$ and 
position space for 5D wavefunction ($\X_n,\XXT_n$).
Imposing the 4D EOMs for the free KK fields as usual,
$\,(\dif_\mu^2+M_{an}^2)V^{a\nu}_n(x) -
   (1-\xi^{-1})\dif^\mu\dif^\nu V^{a}_{n\mu}(x)=0\,$ 
and 
$\,(\dif_\mu^2+\MT_{an}^2)V^{a5}_n(x)=0\,$,\,
we derive from (\ref{eq:5D-EOM}),
\beq
\label{eq:5D-EOM-X}
-\ka^2\dif_5\(h^2\dif_5\X_n\) \,=\, M_{an}^2\X_n \,,~~~~~
-\dif_5(\ka^2\dif_5h^2\XXT_n) \,=\, \xi^{-1}\MT_{an}^2\XXT_n\,,
\eeq
where
\,$(M_{an},\,\MT_{an})$\, are mass-eigenvalues of 
4D KK fields $(V^{a\mu}_n,\,V^{a5}_n)$\,.\,  
Acting $\dif_5$ on the first equation of (\ref{eq:5D-EOM-X}) 
we see, \,$\XXT_n\propto \dif_5 \X_n$\, and
$\,\MT^2_{an}=\xi M^2_{an}\,$.\,
So, with the normalization conditions of 
\,$(\X_n,\XXT_n)$\,,\, we deduce from (\ref{eq:5D-EOM-X}),
\beq
\label{eq:Xn-XTn}  
\dif_5 \X_n \,=\,\dis M_{an}\XXT_n \,,~~~~~~
\dif_5 (h^2\XXT_n) \,=\, -\ka^{-2}M_{an}\X_n \,.
\eeq
With these we decompose the 5D gauge-fixing function
$\Fh^a$ as
\beq
\label{eq:F_an}
\ba{cl}
\Fh^a(\xh ) & \,=\,\dis
\f{1}{\sqrt{L\,}\,}
\sum_{n=0}^{\infty} F^{an}(x)\, \X_n(x^5)/\ka(x^5) \,,
\\[5mm]
F^{an}(x) & \,=~ \dif_\mu V_n^{a\mu}(x) -
                 \xi M_{an} V_n^{a5}(x) 
            \,\equiv\, \Kb_{an}^T \Vnab
\,, 
\ea
\eeq
where $\,\Kb_{an}=(\dif_\mu,\,-\xi M_{an})^T\,$ and
$\,\Vnab =(V_n^{a\mu},\,V^{a5}_n)^T\,$.\,
Thus we derive a 4D ST identity from  
(\ref{eq:5D-ETI1}),
\beq
\label{eq:5D-ETI1-KK}
\lan 0|T 
F^{a_1n_1}(x_1^{~})
F^{a_2n_2}(x_2^{~})
\cdots
F^{a_{\ell}n_{\ell}}(x_{\ell})\Phip |0\ran \,=\, 0\,.
\eeq 
Adding the KK expansions for ghost fields,
\beq
\dis
\ch^a(\xh) = \f{1}{\sqrt{L}\,}
\sum_{n=0}^{\infty} c^a_n(x)\X_n(x^5)\,,~~~~~
\cbh^a(\xh)= \f{1}{\sqrt{L}\,}
\sum_{n=0}^{\infty}\cb^a_n(x)\X_n(x^5)/\ka(x^5)\,,
\eeq
we derive the BRST transformations for the KK fields,
\beq
\label{eq:BRST-4D-KK}
\ba{ll}
\sh V^{a\mu}_n = D^{ab,\mu}_{nm}c^b_m \,,~~~~
&
\sh V^{a5}_n   = D^{ab,5}_{nm}  c^b_m \,,
\\[2mm]
\sh c^a_n      = 
           -\f{1}{2}gC^{abc} 
           \mathfrak{D}_n^{m\ell}c^b_mc^c_{\ell}\,,~~~~
&
\sh \cb^a_n    = -\xi^{-1}F^{an} \,,
\ea
\eeq
where 
\beq
\label{eq:BRST-Dfunc}
\ba{ll}
\dis
D^{ab,\mu}_{nm} & =\dis
-\delta^{ab}\delta_{nm}\dif^\mu 
+\[\f{1}{L}\int_0^L\!\! dx^5\ka^{-2}\X_n\X_m\X_{\ell}\] 
gC^{abc}V_{\ell}^{c\mu}  \,,
\\[4mm]
\dis
D^{ab,5}_{nm} & =\dis
-\delta^{ab}\delta_{nm}M_{am} 
+\[\f{1}{L}\int_0^L\!\! dx^5 h^2\XXT_n\X_m\XXT_{\ell}\] 
gC^{abc}V_{\ell}^{c5}  \,,
\\[4mm]
\dis
\mathfrak{D}_n^{m\ell} & =\dis
\f{1}{L}\int_0^L\!\! dx^5\, \ka^{-2}\X_n\X_m\X_{\ell} \,,
\ea
\eeq
and repeated indices are summed up.
To amputate the external fields $F^{an}$ in 
(\ref{eq:5D-ETI1-KK}),  
we deduce an ST identity for the KK propagator
of \,$V^a_n$, similar to Eq.\,(\ref{eq:ST-D}),
\beq
\label{eq:ST-D-5D}
\Kb_{an}^T\under{\CD}_{nm}^{ab}(p) 
= -[\under{\Omega}_{nm}^{ab}](p)^T   \,,
\eeq
with
\beq
\label{eq:ST-Dx-5D}
\ba{l}
\CD_{nm}^{ab} (p) 
=\lan0|T{\und{V^a_n}}\,\und{V^b_m}^T|0\ran (p) \,,~~~
\Sc_{nm} (p)\delta^{ab} 
=\lan0|Tc^a_n\cb_m^b|0\ran (p) \,,
\\[2mm]
\Omegab_{nm}^{ab}(p) 
\equiv \Omegahb_{mj}^{ab}(p)
                        \Sc_{jn} (p) \,,
\\[3mm]  
\Omegahb_{mj}^{ab} (p) \equiv
\lan 0|T\sh\und{V^b_m}|\cb^a_j\ran (p)
\equiv \!
\(\ba{c} \lan 0|T\sh V_m^{b\mu}|\cb^a_j\ran
\\[1mm]  \lan 0|T\sh V_m^{b5}|\cb^a_j\ran 
\ea
\)_{\!\!\!(p)}\!\! \equiv\!
\(\!\ba{c} 
-ip^\mu\delta^{ab} 
    [\d_{jm}\!+\!\Delta^a_{jm}(p^2)]
\\[1mm]  
-M_{am}\delta^{ab} 
    [\d_{jm}\!+\!\DeT^a_{jm}(p^2)] 
\ea
\)\!.
\ea 
\eeq
Extension of the above formulation to include possible brane
term is straightforward. Consider, for simplicity, a brane term
similar to (\ref{eq:MooseB-BK})-(\ref{eq:MooseB-BKx}) in Sec.\,4,
at $\,y=0,\,{\rm or}~L$,\, 
\beq
\label{eq:BT0}
\! S_{\rm BT}  = \!\!
\dis\int\!\! d^5\xh ~\delta(x^5\!\!-\!y)\,
\f{-\zeta\,}{4} \Fh^{a\mu\nu}\Fh^a_{\mu\nu} 
               = \!\!
\dis\int\!\! d^5\xh~
\delta (x^5\!\!-\!y)\sqrt{-\gh\,}\,\gh^{IK}\gh^{JL}
\f{-\zetah\,}{\,4\ka^2} \Fh^a_{IJ}\Fh^a_{KL}\,,\, 
\eeq
where $\,\zetah =\zeta\, \ka^2(y)\,$.
In (\ref{eq:BT0}) the second expression is 
equivalent to the first one under the covariant
BC $\Fh^a_{\mu 5}=0$ [cf. (\ref{eq:5D-covBC})], and appears
more convenient for direct extension of the above derivation.
The brane term (\ref{eq:BT0}) may vary, depending
on the residual symmetry at the boundary.
We stress that the total action $\,S_5+S_{\rm BT}\,$
is invariant under the 5D gauge symmetry supplemented by the
BCs [which are {\it induced from the continuum limit of the 
gauge-invariant moose theory (\ref{eq:L})}, cf. Sec.\,4].
The quantized action may be written as      
$\,S_5+S_{\rm BT}+S_{\rm gf}+S_{\rm FP}\,$ with 
$\,S_{\rm gf}+S_{\rm FP}
   =\int\! d^5\xh \,[1+\zetah\delta(x^5\!-y)]
    (\LLH_{\rm gf}+\LLH_{\rm FP})\,$,\,    
which is expected from the deconstruction viewpoint since
the gauge group at each lattice site of the 5D 
(including boundary sites)
can be quantized with its own gauge-fixing and ghost terms as shown
earlier in this section.
With these and a prescription of the vanishing total derivative
$\,\dif_5(\cdots)\,$ under the integration 
$\,\int\! dx^5[1+\zetah\delta(x^5-y)]\,$,\,
we readily verify that the above 
Eqs.\,(\ref{eq:BRST-5D})-(\ref{eq:ST-Dx-5D}) remain,
except that all integrals in (\ref{eq:BRST-Dfunc}) will contain
an extra brane-term-factor \,$[1+\zetah\delta(x^5-y)]$.\, 
Accordingly, the 5D wavefunctions $(\X_n,\,\XXT_n)$ satisfy
the normalization conditions,
\beq
\label{eq:NCkin}
\ba{rcl}
L^{-1}\!\int\!dx^5[1+\zetah\delta(x^5\!-\!y)]
    \ka^{-2}\X_n\X_m &~=~& \delta_{nm} \,,
\\[2mm]
L^{-1}\!\int\!dx^5[1+\zetah\delta(x^5\!-\!y)]h^2
    \XXT_n\XXT_m     &~=~& \delta_{nm} \,,
\ea
\eeq 
on the general 5D background.
The corresponding normalization conditions for 
the KK mass-terms are,
\beq
\label{eq:NCmass}
\ba{rcl}
  L^{-1}\!\int\!dx^5[1+\zetah\delta(x^5\!-\!y)]
      h^2\dif_5\X_n\dif_5\X_m 
&~=~& M_{an}^2\delta_{nm} \,,
\\[2mm]
 L^{-1}\!\int\!dx^5[1+\zetah\delta(x^5\!-\!y)]\ka^2
    \dif_5(h^2\XXT_n)\dif_5(h^2\XXT_m) 
&~=~& M_{an}^2\delta_{nm} \,. 
\ea
\eeq
Under Eq.\,(\ref{eq:Xn-XTn}) we see that (\ref{eq:NCmass})
consistently reduces to (\ref{eq:NCkin}).

With these given, we are ready to amputate the external fields in
(\ref{eq:5D-ETI1-KK}) and derive an identity 
for $S$-matrix elements,
\beq
\label{eq:5D-ETI2}
\T\[\FB^{a_1n_1}\!(p_1^{~}),\,
    \FB^{a_2n_2}\!(p_2^{~}),\,\cdots,\,
    \FB^{a_{\ell}n_{\ell}}\!(p_{\ell}^{~}),\Phip\] 
~=~0 \,,
\eeq
where 
$\,\FB^{an}=  p^\mu V^{an}_{\mu} 
             -M_{an} \Ch^a_{nm} V_5^{am}
           =  M_{an} (V^{an}_S-\Ch^a_{nm} V_5^{am})
\,$,\,  $V_S^{an}=\ep_S^{\mu}V_{\mu}^{an}$\,
[$\ep_S^{\mu}\equiv k^{\mu}/M_{an}
  =\ep_L^{\mu}-v^{\mu}$,\,
 $v^{\mu}=O(M_{an}/E_n)$\,],\,
and 
$\,\Ch^a_{nm} \equiv i(M_{am}/M_{an})  
  [(\mathbf{1}+\DeT^a(p^2))
   (\mathbf{1}+\D^a(p^2))^{-1}]_{mn}$\,.\,
Similar to (\ref{eq:4D-ETI2}) for the deconstruction theory,     
our ET identity (\ref{eq:5D-ETI2}) shows that 
the unphysical scalar-KK-component $V^{an}_S$ and the fifth
gauge-KK-component $V^{an}_5$ (or its linear composition
$\Ch^a_{nm}V_5^{am}$) are {\it confined} at the $S$-matrix level,
so they together have zero contribution to any physical process.
This is just the quantitative $S$-matrix formulation of the 5D 
{\it geometric Higgs mechanism} (GHM)\cite{SUSY03,CDH},
where $V^{an}_5$\,'s serve as the
would-be Goldstone bosons and get converted to the 
longitudinal gauge KK-modes $V_L^{an}$\,'s.  
Expanding the ET identity (\ref{eq:5D-ETI2}), 
we thus derive a {\it general Kaluza-Klein ET}
(KK-ET\cite{CDH}) at asymptotic energy,  
\beq
\label{eq:KK-ET}
\dis
\T\!\[V_L^{a_1^{~}n_1^{~}},\,
      V_L^{a_2^{~}n_2^{~}},
\cdots\,\] ~=~ 
\Ch_{\rm mod}^{n_1m_1,n_2m_2\cdots}\,
\T\!\[V_5^{a_1^{~}m_1^{~}},\,
      V_5^{a_2^{~}m_2^{~}},
      \cdots\,\] 
      \,+\, O\!\(\!\f{M_{an}}{E_n}\!\) \!,
\eeq
where 
\,$\Ch_{\rm mod}^{n_1m_1,n_2m_2\cdots} 
   =\Ch^{a_1}_{n_1m_1}\cdots 
    \Ch^{a_{\ell}}_{n_{\ell}m_{\ell}}
   = i^{\ell}
    [(\delta_{n_1m_1}\delta_{n_2m_2}\cdots )
     +O({\rm loop})]$\,,\,
and the external momenta are put on-shell
($p_n^2=M_{an}^2$).\, 
The exact expression of the suppressed term 
$\,O(M_{an}/E_n)\,$ in (\ref{eq:KK-ET})
is tedious and is directly obtained from 
expanding the ET identity (\ref{eq:5D-ETI2}) by
\,$\ep_S^{\mu}= \ep_L^{\mu}-v^{\mu}$,\,
in the {\it same} way as in [\refcite{ET}].\footnote{As
well-known\cite{ET,CDH,SUSY03}, the exact ET identity
(\ref{eq:5D-ETI2}) ensures the sum of the
Goldstone amplitude and the lengthy $O(M_{an}/E_n)$ term to  
precisely equal the amplitude of longitudinal gauge bosons.}
We stress that {\it our KK-ET (\ref{eq:KK-ET}) 
and its KK-ET identity (\ref{eq:5D-ETI2})
are valid for arbitrary geometry with any consistent BC 
and possible brane term such as (\ref{eq:BT0})
(cf. Sec.\,4 for detail).}\footnote{After the
completion of this manuscript, a preprint\cite{muck}
appeared which considered a generalization of the 5D ET  
in Refs.\,[\refcite{CDH,SUSY03}] to include brane term 
in the special case of flat $S^1/Z_2$.}
The KK-ET (\ref{eq:KK-ET}) is the manifestation of the 5D
geometric Higgs mechanism\cite{SUSY03,CDH} at
the $S$-matrix level, where 
the conversion $\,V^{an}_5\Longrightarrow V^{an}_L\,$\,
is realized. 
An essential advantage of this 5D formulation is that we
start with the well-defined 4D moose theory (gauged nonlinear 
sigma model) which is manifestly gauge-invariant without any
BC.   The consistent BCs and possible brane terms  
are {\it automatically induced} 
by taking proper continuum limits (cf. Sec.\,4),
which also ensure the gauge-invariance of the
5D action at the boundaries.

The gauge interactions of $V_5^{an}$ arise from 
$\,\int d^5\xh \f{-h^2}{2}\Fh^a_{\mu 5}\Fh^{a\mu 5}\,$
under KK expansion.
Examining this\cite{CDH,SUSY03} and
applying power counting\cite{PC},
we find that scalar amplitude
$\,\T[V_5^{a_1^{~}n_1^{~}},\,V_5^{a_2^{~}n_2^{~}},
         \cdots\cdots ]=O(E^0)$\, 
generally holds\cite{SUSY03},\,
{\it independent of the detail of
any particular compactification/BC and 
possible brane term.} 
Because of this and Eq.\,(\ref{eq:KK-ET}),
we conclude that the longitudinal amplitude
$\,\T[V_L^{a_1^{~}n_1^{~}},\,
      V_L^{a_2^{~}n_2^{~}},\cdots\cdots ]=O(E^0)\,$,\,
generally guaranteeing the $E$-power cancellations down to  
a constant.

Finally, we note that the nonlinearly realized gauge symmetry 
in the moose theories
may be regarded as a (spontaneously) broken phase formulation, 
but the crucial difference from
an arbitrary or random {\it explicit symmetry breaking} is that 
our formalism (including the continuum 5D action) 
is manifestly gauge-invariant or BRST-invariant 
in which the would-be Goldstone bosons
and the corresponding ``Higgs'' mechanism are consistently 
embedded, so the various Slavnov-Taylor and Ward-Takahashi 
identities can be derived even in the broken phase\cite{BWLee}.\,
It is such {\it general} identities and their resulting 
ET (\ref{eq:ETm}) or KK-ET (\ref{eq:KK-ET}) 
that have directly guaranteed the relevant $E$-cancellations 
in {\it all} \,\,$2\to n$ ($n\geqq 2$)\,\, 
longitudinal gauge boson scatterings for {\it any} 
4D or compactified higher-D theory, 
leading to the effective unitarity.\cite{CH,CDH,SUSY03}

\section{Deconstruction, (Non)Geometric Gauge Symmetry Breaking 
         and Induced Boundary Conditions}

Various compactified 5D theories can be derived by taking 
proper continuum limit of our general moose theory (Fig.\,1).
The bulk gauge symmetry and its breaking at the boundary 
are essentially determined by the structure of 4D moose 
and its spontaneous symmetry breaking pattern.
In comparison with the traditional 4D Higgs mechanism\cite{Higgs} 
which spontaneously breaks the gauge symmetry by nontrivial 
vacuum, the 5D gauge symmetry can be spontaneously broken
by the proper compactification at boundaries which arise
from the ``boundary'' groups (sites) of the moose theory
(Fig.\,1) under possible folding(s). 
This provides us a
conceptually clean and elegant formulation of the gauge
symmetry breaking (without/with gauge-group rank reduction) 
in compactified 5D theories, 
without inputting the BCs {\it a priori} or
relying on the technique of adding 
extra boundary Higgs fields\cite{Heb}.\,
Our generic moose analysis below also reveals that the
induced 5D BCs for $\Ah^{a\mu}$ do not 
depend on particular choice of bulk geometry,
while the same is not generally true for $\Ah^{a5}$.

\begin{figure}[h]
\label{fig:Fig2}
\begin{center}
\vspace*{-5mm}
\centerline{\psfig{file=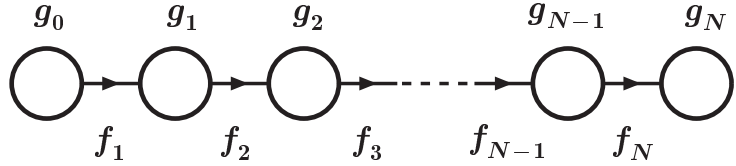,width=11cm}}
\setlength{\unitlength}{1mm}
\begin{picture}(0,0)
\end{picture}
\vspace*{-8mm}
\caption{Higgsless Moose-A: a special case of Fig.\,1
with \,$M=0$\, and \,$f_{N+1}^{~}=0$\,.\, 
The gauge symmetry breaking pattern is
$\,SU(2)^{N+1}\to SU(2)_D\,$, 
with massless zero modes.}
\end{center}
\vspace*{-5mm}
\end{figure}

We start by considering the special case of Fig.\,1 with 
\,$M=0$\, and \,$f_{N+1}^{~}=0$\,,\, whose gauge symmetry
breaking pattern is $\,SU(2)^{N+1}\to SU(2)_D\,$,\,
as shown in Fig.\,2.  
The $(N+1)\times (N+1)$ gauge boson mass-matrix 
$\MMB_a^2 =\MMB_{[0,N]}^2$ 
is given in (\ref{eq:M_CC}) with \,$f_{N+1}^{~}=0$\,,\, 
and its orthonormal eigenvectors are denoted as 
\,$\XXB_n = (X_{0n},\,X_{1n},\,\cdots,\,X_{Nn})^T$\,
($n=0,1,2,\cdots,N$).   So the gauge-eigenbasis field
$\AAB^{a\mu}$ and mass-eigenbasis field 
$\VVB^{a\mu}$ are connected
by   $\,\AAB^{a\mu} = \RRB_a^{~}\VVB^{a\mu}\,$
with rotation
$\,\RRB_a^{~} = \(\XXB_0,\,\XXB_1,\,\cdots,\,\XXB_N\)\,$,\,
i.e.,
\beq
\label{eq:X-X5D}
\dis
A^{a\mu}_j ~=
\sum_{n=0}^N V_n^{a\mu} X_{jn} \,,
~~~\Longrightarrow~~~
\Ah^{a\mu}(\xh ) ~= \dis \f{1}{\sqrt{L\,}\,}
\sum_{n=0}^{\infty} V_n^{a\mu}(x) \X_{n}(x^5) \,,
\eeq
where the second relation is 
the corresponding continuum KK-expansion
under \,$N\to \infty$\, and $\,j\,{\bf a}\to x^5\,$.\,
Analyzing the eigenvalue equation 
$\,\MMB_a^2 \XXB_n =\lambda_n \XXB_n\,$,\,
we find that the eigenvectors satisfy the
following consistency relations,
\beq
\label{eq:4D-BC0}
\dis
g_0^{~}X_{0,n} - g_{-1}^{~} X_{-1,n} \,=\, 0\,, ~~~~~
g_{N+1}^{~} X_{N+1,n} - g_N^{~} X_{N,n} \,=\, 0\,.
\eeq
As shown above Eq.\,(\ref{eq:SU2-5D-L}), taking the
5D continuum limit leads to the identification,
 $\, A^{a\mu}_j(x) \to
     (g_5^{~}/g_j^{~})\Ah^{a\mu}(x,x^5_j)
 $\,.\, 
Combining this with (\ref{eq:X-X5D}) we find that the eigenvector
$\,X_{jn}\,$ is related to the 5D KK-wavefunction $\X_n(x^5)$ 
via $\,g_j^{~}X_{jn}\to (g_5^{~}/\sqrt{L}\,)\X_n(x_j^5)$\,.\, 
Hence, we rewrite (\ref{eq:4D-BC0}) as conditions for the
5D KK-wavefunction $\,\X_n(x_j^5)\,$,
\beq
\label{eq:5D-BC0j}
\dis
\X_{n}(x_0^5)     - \X_{n}(x_{-1}^5) \,=\, 0\,, ~~~~~
\X_{n}(x_{N+1}^5) - \X_{n}(x_N^5)    \,=\, 0\,.
\eeq
From (\ref{eq:5D-BC0j}),
we thus derive the {\it induced boundary conditions (BCs)} 
of Neumann type in the 5D continuum limit,
\beq
\label{eq:5D-BC0}
\left.
\dif_5^{~}\Ah^{a}_\mu \right|_{x^5=0} \,=\,0\,,
~~~~~~
\left.
\dif_5^{~}\Ah^{a}_\mu \right|_{x^5=L} \,=\,0\,,
\eeq
where $\,L=N\aa\,$ is the length of 5D.
This compactified 5D theory has an unbroken
$SU(2)$ gauge symmetry survived in 4D (corresponding to the 
residual diagonal group $SU(2)_D$ of the Moose-A in Fig.\,2), 
which ensures the massless zero KK-modes. 
So, the Neumann BCs (\ref{eq:5D-BC0}) 
break a 5D $SU(2)$ into a 4D $SU(2)$
(with massive gauge KK towers as well as massless zero modes), 
and preserve the rank of 5D gauge group, precisely matching
the symmetry breaking structure of its parent Moose-A
in Fig.\,2.

We can further derive the BCs for $\Ah^{a5}$. In the continuum
limit the site-Goldstone fields $\pi^a_j$ will geometrize as $\Ah^{a5}$
[cf. (\ref{eq:U-lattice})], 
while the eaten Goldstones $\pit^a_n$ will geometrize
as the KK modes $V^{a5}_n$
[cf. (\ref{eq:pi-A5-geo}) below].
From the mass-diagonalization in Eqs.\,(\ref{eq:M2-diag}) and
(\ref{eq:PI-rotation}) and 
the $R_\xi$ gauge-fixing in Eq.\,(\ref{eq:GF}),
we see that the eaten Goldstone fields $\Pit^a$ have mass matrix
$\xi_a(\MMBT_a^2)^{\rm diag}$ and the site-Goldstones $\Pi^a$ have
mass matrix $\xi_a\MMBT_a^2$.\,  For the Moose-A (Fig.\,2), we
derive the $N\times N$ mass matrix $\MMBT_a^2$,
{\small
\begin{equation}
\label{eq:MA-dual}
\MMBT_a^{2} = \df{1}{4}
\left(
\begin{array}{c|c|c|c|c}
&&&&
\\[-2mm]
(g_0^2+g_1^2) f_1^2    & -g_1^2f_1^{~}f_2^{~}     &      &
\\[1.5mm]
\hline
&&&&
\\[-2mm]
   -g_1^2f_1^{~}f_2^{~}
&  (g_1^2\!+\!g_2^2)f_2^2
&  -g_2^2f_2^{~}f_3^{~}         &
\\[1.5mm]
\hline
&&&&
\\[-2mm]
& -g_2^2f_2^{~}f_3^{~}     
& (g_2^2 \!+\! g_3^2)f_3^2 
& -g_3^2f_3^{~}f_4^{~}   
\\[1.5mm]
\hline
&&&&
\\[-4mm]
 & & \ddots & \ddots & \ddots
\\[0.5mm]
\hline
&&&&
\\[-2mm]
 & & &  -g_{N\!-\!1}^2 f_{N\!-\!1}^{~}f_N^{~}    &
        (g_{N\!-\!1}^2\!+\!g_N^2)f_N^2               
\\[1.5mm]
\end{array}
\right)\!,~
\end{equation}
}
\noindent\hspace*{-2mm}      
whose orthonormal eigenvectors are denoted by 
$\,\XXBT_n=(\XT_{1n},\,\XT_{2n},\,\cdots,\,\XT_{Nn})^T$\,
($n=1,2,\cdots,N$).\,   
The orthogonal rotation matrix in (\ref{eq:PI-rotation})
is thus given by
$\,\RRB_a^{~} = (\XXBT_1,\,\XXBT_2,\,\cdots,\,\XXBT_N)\,$.\,
So, in the continuum limit we have the following geometrization,
\beq
\label{eq:pi-A5-geo}
\dis
\pi^{a}_j ~=
\sum_{n=0}^N \pit_n^a \XT_{jn} \,,
~~~\Longrightarrow~~~
\Ah^{a5}(\xh ) ~= \dis
\f{1}{\sqrt{L\,}\,}
\sum_{n=0}^{\infty} V_n^{a5}(x) \XXT_{n}(x^5) \,.
\eeq
As shown below Eq.\,(\ref{eq:U-lattice}), the site-Goldstone field 
$\,\pi^a_j = (g_j^{~}f_j^{~}\aa/2)A^a_{5}(x,x^5_j)\,$ 
is connected to the 5D gauge field $\,\Ah^a_5(x,x_j^5)\,$ via
$\,\pi^a_j \to (\aa g_5^{~}f_j^{~}/2)\Ah^a_5(x,x_j^5) 
$\, in the continuum limit.
So, from (\ref{eq:pi-A5-geo}) we find that $\XT_{jn}$ and
$\XXT_n(x^5_j)$ are connected via
$\,\XT_{jn}\to (\aa g_5^{~}f_j^{~}/2\sqrt{L})\XXT_n(x_j^5)
\,$.\,
Analyzing the eigenvalue equation 
$\,\MMBT_a^2\XT_n = \lambda_n \XT_n\,$,\,
we find that the eigenvectors must satisfy the following
consistency conditions,
\beq
\label{eq:MooseA-XTn-BC}
\XT_{0,n}=\,0\,, ~~~~~
\XT_{N+1,n}=\,0\,; ~~~\Longrightarrow~~~
\XXT_{n}(x_0^5)\,=\,0\,, ~~~~~
\XXT_{n}(x_{N+1}^5)\,=\,0\,.
\eeq
In the continuum limit, they result in the {\it induced} 5D BCs
for $\Ah^a_5$, 
\beq 
\label{eq:5D-BC0-A5}
\left.\Ah^a_5\right|_{x^5=0}\,=~0\,, ~~~~~~
\left.\Ah^a_5\right|_{x^5=L}\,=~0\,,
\eeq
which are both of Dirichlet type, unlike the 
Neumann BCs for $\Ah^{a\mu}$ in (\ref{eq:5D-BC0}).
This is fully consistent with  
the relation $\,\dif_5\X_n \propto \XXT_n\,$
in (\ref{eq:Xn-XTn}) for 5D KK-wavefunctions. 
We also note that the Neumann BCs (\ref{eq:5D-BC0})
for $\Ah^{a\mu}$ and the Dirichlet BCs (\ref{eq:5D-BC0-A5})
for $\Ah^{a5}$ can be combined into a gauge-covariant form,
\beq
\label{eq:5D-covBC}
\left.\Fh^a_{\mu 5}\right|_{x^5=0}\,=~0\,, ~~~~~~
\left.\Fh^a_{\mu 5}\right|_{x^5=L}\,=~0\,,
\eeq
which explicitly retains the gauge-invariance of the 5D action
at the corresponding boundary.

\vspace*{3mm}
We then turn to a different gauge symmetry
breaking structure. This is most cleanly described by
the Moose-B in Fig.\,3 (which we called ``enlarged moose''
in 2001\cite{SUSY03}).
This is a special case of our general moose theory in
Fig.\,1 with \,$M=0$\, and \,$g_{N+1}^{~}=0$\,,\, 
with gauge group breaking pattern 
$\,SU(2)^{N+1}\!\to\,$nothing\,.\, 
As we will show, it {\it induces} the 
Dirichlet type BCs in 5D continuum limit and
reduces the rank of the 5D gauge group.

\begin{figure}[h]
\label{fig:Fig3}
\begin{center}
\vspace*{-7mm}
\centerline{\psfig{file=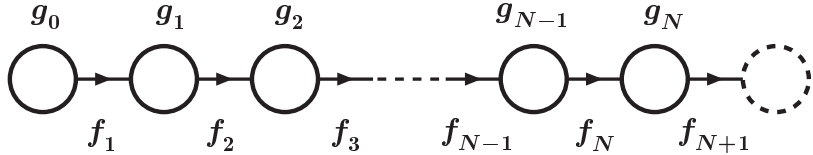,width=12cm}}
\setlength{\unitlength}{1mm}
\begin{picture}(0,0)
\end{picture}
\vspace*{-8mm}
\caption{Higgsless Moose-B: a special case of Fig.\,1
with \,$M=0$\, and \,$g_{N+1}^{~}=0$\,.\, 
The gauge symmetry breaking pattern is
$\,SU(2)^{N+1}\!\to\,$\,nothing, 
without massless ``zero modes''.}
\end{center}
\vspace*{-5mm}
\end{figure}

The gauge boson mass matrix 
$\,\MMB_a^2=\MMB_{[0,N+1)}^2\,$
is just given by
Eq.\,(\ref{eq:M_CC}), 
which has massive eigenvalues only.
Analyzing the eigenvalue equation 
$\,\MMB_a \XXB_n =\lambda_n \XXB_n\,$,\,
we observe that the eigenvectors satisfy the
following consistency relations,
\beq
\label{eq:4D-BC1}
\ba{lll}
&  g_0^{~}X_{0,n} - g_{-1}^{~}X_{-1,n} \,=\, 0\,, ~~~&~~~
   X_{N+1,n} \,=\, 0\,;
\\[2.5mm]
\Longrightarrow~~ &
\X_{n}(x_0^5) - \X_{n}(x_{-1}^5) \,=\, 0\,, ~~~&~~~
\X_{n}(x_{N+1}^5)                \,=\, 0\,.
\ea
\eeq
Taking the continuum limit as before, we derive the induced
BCs, 
\beq
\label{eq:5D-BC1}
\left.
\dif_5^{~}\Ah^{a}_\mu \right|_{x^5=0} \,=\,0\,,
~~~~~~~~
\left.
\Ah^{a}_\mu \right|_{x^5=L} \,=\,0\,,
\eeq
where we see that similar to (\ref{eq:5D-BC0}) 
the BC at $x^5=0$ remains the Neumann type, but
the new condition at $x^5=L$ is a Dirichlet BC.
The moose representation of Fig.\,3 clearly shows
that it is this induced Dirichlet BC at $x^5=L$
that completely breaks the 5D $SU(2)$ gauge group 
and makes all zero KK-modes massive.
The above formulation based on the Moose-A and -B 
can be directly generalized to any
5D gauge theory with an arbitrary 
simple group \,$G_A$\,.

We have also derived the BCs for $\Ah^{a5}$ 
in the Moose-B.  Consider the mass matrix
$\xi_a\MMBT_a^2$ for site-Goldstone fields $\Pi^a$,
where the $(N+1)\times (N+1)$ matrix $\MMBT_a^2$ 
can be obtained from (\ref{eq:MA-dual}) by simple 
replacements $\,N\to N+1\,$ and \,$g_{N+1}^{~}\to 0$\,.\,
Analyzing the eigenvalue equation
$\,\MMBT_a^2\XXBT_n=\lambda_n\XXBT_n\,$, we deduce the
following consistency relations for the eigenvectors,
\beq
\label{eq:MooseB-XTn-BC}
\XT_{0,n} \,=~0\,,~~~~~~~~
f_{N+2}^{~}\XT_{N+2,n} - f_{N+1}^{~}\XT_{N+1,n} \,=~0\,.
\eeq
As before, we have the identification, 
$\,\XT_{jn}\to (\aa g_5^{~}f_j^{~}/2\sqrt{L})\XXT_n(x_j^5)
\,$,\, in the continuum limit.
So we can rewrite (\ref{eq:MooseB-XTn-BC}) as conditions
for $\XXT_n(x_j^5)$,
\beq
\label{eq:5D-BC1-A5j}
\XXT_{n}(x_0^5) \,=~0\,,~~~~~~~~
f_{N+2}^{2}\XXT_n(x_{N+2}^5) - 
f_{N+1}^{2}\XXT_n(x_{N+1}^5) \,=~0\,.
\eeq
With the definition $\,f_j^{~}=h_jf=h(x^5_j)f\,$,\, 
we derive the {\it induced BCs} 
from (\ref{eq:MooseB-XTn-BC})
in the 5D continuum limit,
\beq
\label{eq:5D-BC1-A5}
\left.
\Ah^{a}_5 \right|_{x^5=0} \,=\,0    \,,
~~~~~~~~
\left.
\dif_5 (h^2\Ah^a_5 ) \right|_{x^5=L} \,=\,0 \,.
\eeq
They are fully consistent with the BCs for $\Ah^a_{\mu}$
in (\ref{eq:5D-BC1}) according to Eq.\,(\ref{eq:Xn-XTn})
which gives 
$\,\dif_5\X_n        \propto\XXT_n\,$ and 
$\,\dif_5(h^2\XXT_n) \propto\X_n\,$.
Note that, unlike the Dirichlet BC for $\Ah^a_5$,
the Neumann BC for $\Ah^a_5$ does depend on the bulk
geometry via the function \,$h(x^5)$\,.
Implications of this will be 
explored elsewhere\cite{HLmoosenew}.

\begin{figure}[h]
\label{fig:Fig4}
\begin{center}
\vspace*{-3mm}
\centerline{\psfig{file=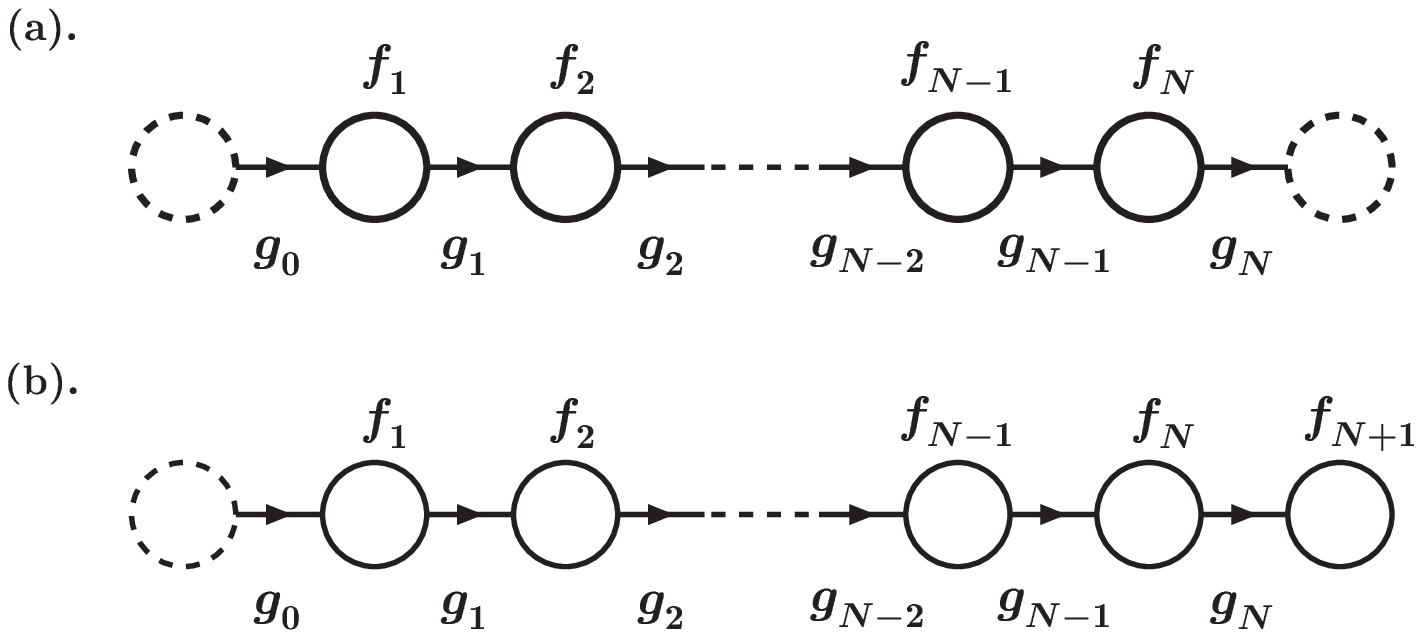,width=12cm}}
\setlength{\unitlength}{1mm}
\begin{picture}(0,0)
\end{picture}
\vspace*{-5mm}
\caption{Two dual mooses: 
(a). Moose-$\rm\At$ is the dual of Moose-A in Fig.\,2;
(b). Moose-$\rm\Bt$ is the dual of Moose-B in Fig.\,3.
}
\end{center}
\vspace*{-5mm}
\end{figure}

It is also instructive to view the Goldstone mass matrix
$\MMBT_a^2$ as the ``gauge boson'' mass matrix generated 
in the corresponding dual moose\cite{dualm,HLmoose}
which is defined by the exchange 
\,$\{g_n^{~}\}\longleftrightarrow
   \{f_n^{~}\}$\, from the original moose.
We show the dual versions of the Moose-A and -B in Fig.\,4.
With the notations of open/closed intervals, it may be
convenient to denote the dual mass matrix as 
\,$\MMBT_a^2=\MMBT_{(0,N+1)}^2$\, for Moose-${\rm\At}$ 
(where the sites \,$j=0,\,N\!+\!1$\, themselves
have no contribution due to $\,f_0^{~}=f_{N+1}^{~}=0$\,), 
and  \,$\MMBT_a^2=\MMBT_{(0,N+1]}^2$\, 
for Moose-${\rm\Bt}$ (where the site $j=0$ has 
no contribution due to $\,f_0^{~}=0$\,).\,
Comparing the dual Moose-${\rm\At}$ [Fig.\,4(a)] with 
Moose-A [Fig.\,2], we can now intuitively see why the
BCs for $\XXBT_n$ and $\Ah^{a5}$ in 
(\ref{eq:MooseA-XTn-BC})-(\ref{eq:5D-BC0-A5})
just take Dirichlet type, different from the Neumann BCs 
(\ref{eq:4D-BC0})-(\ref{eq:5D-BC0})
for $\XXB_n$ and $\Ah^{a\mu}$.\,
Similarly, comparison of the dual Moose-$\rm\Bt$ [Fig.\,4(b)] 
with Moose-B [Fig.\,3] explains the difference between the 
BCs (\ref{eq:5D-BC1-A5j})-(\ref{eq:5D-BC1-A5}) and 
(\ref{eq:4D-BC1})-(\ref{eq:5D-BC1}).

\vspace*{2mm}
Another advantage of our general deconstruction
formalism is that we can also derive the proper
brane kinetic terms in the continuum limit.
Considering our Moose-B for instance, we quantify
how a brane term at \,$x^5=0$\,   
is reconstructed from localizing
the boundary group \,$SU(2)_0$\, at site \,$j=0$\,.\,  
Setting $\,f_1^2\gg f_{j}^2\,$ ($j\geqq 2$) 
for taking the continuum limit 
(which prevents the site-$0$ from 
joining the 5D geometry), we examine the equation
of motion (EOM) for $A^{a\mu}_0$ and derive the following
relation,
\beq
\label{eq:W3_0}
A_{0}^{a\mu} ~=~ 
\dis\({g_1^{~}}/{g_{0}^{~}}\)A_1^{a\mu} \,.
\eeq
In the continuum theory,
this localizes \,$SU(2)_0$\, as a brane kinetic term at
$\,x^5=0\,$,
\beq
\label{eq:MooseB-BK}
S_{\rm BT}^{~} ~= \dis\int\!\! d^5\xh~
\delta (x^5\!-0)\,
\f{\,-(\gh^{~}\!\!/g_{0}^{~})^2\,}{4}
\Fh^a_{\mu\nu}\Fh^{a\mu\nu} \,,
\eeq
where \,$\gh \equiv g_5^{~}$ denotes the 5D gauge coupling.\,
The above brane term may be viewed as being actually localized at 
$\,x^5=0^+$,\, the {\it right-neighborhood} of the point
\,$x^5=0$\, at which our BC 
\,$\,\dif_5^{~}\Ah^{a}_\mu |_{x^5=0} \!=\,0\,$\, 
is imposed. This is because our
deconstruction procedure of deriving this BC from (\ref{eq:4D-BC1}) 
{\it includes} the boundary site \,$j=0$\, itself
plus its left-neighborhood.
We also note that the theory obeys a covariant BC 
$\,\Fh^a_{5\mu}=0\,$ at $x^5=0$ [similar to
(\ref{eq:5D-covBC})], 
so we may rewrite the brane term (\ref{eq:MooseB-BK}) 
to have explicit invariance under the 5D gauge group 
at $x^5=0$,
\beq
\label{eq:MooseB-BKx}
S_{\rm BT} ~= \dis\int\!\! d^5\xh~
\delta (x^5\!-0)\,\sqrt{-\gh\,}\,\gh^{IK}\gh^{JL}
\f{\,-\zetah\,}{\,4\ka^2\,}
\Fh^a_{IJ}\Fh^a_{KL} \,,
\eeq
where $\,\zetah = (\gh/g_0)^2\ka^2 (0)\,$.  
Other brane terms may be rewritten in the similar way.

\vspace*{3mm}
Next, we analyze the Moose-C in Fig.\,5, which
is a special case of Fig.\,1 with 
\,$M=0$\, and \,$g_{N+1}^{~}\neq 0$\,.\,
As a slight variation of Moose-B, 
Moose-C has gauged its subgroup $U(1)_{N+1}$ 
and has the symmetry breaking pattern 
$\,SU(2)^{N+1}\ot U(1)\to U(1)_{\rm em}\,$,\,
as a straightforward extension of  
Ref.\,[\refcite{GBESS}] which studied 
the nontrivial minimal models 
\,$SU(2)^3\ot U(1)\to U(1)_{\rm em}$
($N+1= 3$) and 
\,$SU(2)^2\ot U(1)^2\to U(1)_{\rm em}$
($N+1=M+1= 2$)\, with general inputs
\,$(g_j^{~},\,f_k^{~})$\,.\,

\begin{figure}[h]
\label{fig:Fig5}
\begin{center}
\vspace*{-6mm}
\centerline{\psfig{file=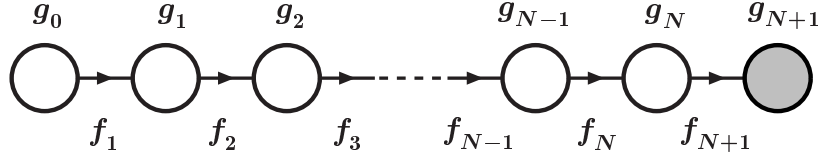,width=12cm}}
\setlength{\unitlength}{1mm}
\begin{picture}(0,0)
\end{picture}
\vspace*{-6mm}
\caption{Higgsless Moose-C: a special case of Fig.\,1
with \,$M=0$\, and \,$g_{N+1}^{~}\neq 0$\,.\, 
The gauge symmetry breaking pattern is
$\,SU(2)^{N+1}\otimes U(1)\to U(1)_{\rm em}\,$,\,
extended from Ref.\,[22].} 
\end{center}
\vspace*{-5mm}
\end{figure}

The charged gauge boson mass matrix 
$\MMW =\MMB^2_{[0,N+1)}$
is given in (\ref{eq:M_CC}), the same as in
Moose-B, while the neutral gauge boson mass matrix
$\MMZ =\MMB_{[0,N+1]}^2$ can be obtained from
$\MMZ$[Moose-A] with a replacement $N\to N+1$.
In consequence we find that the consistency
relations for the eigenvectors 
$\{\XXB^\pm_n\}$ and $\{\XXB^3_n\}$
are just a mixture
of Moose-A and -B in (\ref{eq:4D-BC0}) and
(\ref{eq:4D-BC1}),
\beq
\label{eq:4D-BC2}
\ba{ll}
g_0^{~}X_{0,n}^\pm - g_{-1}^{~}X_{-1,n}^\pm \,=\, 0\,, ~~~&~~
X_{N+1,n}^\pm \,=\, 0\,;
\\[2mm]
g_0^{~}X_{0,n}^3 - g_{-1}^{~}X_{-1,n}^3 \,=\, 0\,, ~~~&~~
g_{N+2}^{~}X_{N+2,n}^3 - g_{N+1}^{~}X_{N+1,n}^3 \,=\, 0\,.
\ea
\eeq
So, with the same reasoning as before, 
we obtain the induced continuum BCs
which combine (\ref{eq:5D-BC1}) for charged sector
and (\ref{eq:5D-BC0}) for neutral sector,
\beq
\label{eq:5D-BC2}
\ba{l}
\left.\dif_5^{~}\Ah^{\pm}_\mu \right|_{x^5=0} \!=\,0\,,
~~~
\left.\dif_5^{~}\Ah^{3}_\mu \right|_{x^5=0} \!=\,0\,;
~~~
\left.\Ah^{\pm}_\mu \right|_{x^5=L} \!=\,0\,,
~~~
\left.\dif_5^{~}\Ah^{3}_\mu \right|_{x^5=L} \!=\,0\,.
\ea
\eeq
In the continuum limit, the boundary group
$U(1)_{N+1}$ will be localized at \,$x^5=L\,$,\,
similar to the brane term which we reconstructed in
(\ref{eq:MooseB-BK}). 
So, we can localize $U(1)_{N+1}$ 
by setting $\,f_{N+1}^{2}\gg f_{j}^{2}\,$  ($j\leqq N$)\,
for the continuum limit.  We quantify this by
examining the EOM for $A_{N+1}^{3\mu}$
under large $f_{N+1}^{2}$ limit, thus we derive a relation,
\beq
\label{eq:W3_N+1}
A_{N+1}^{3\mu} ~=~ 
\dis\({g_N^{~}}/{g_{N+1}^{~}}\)A_N^{3\mu} \,,
\eeq
which generates the localized brane kinetic term at
\,$x^5=L$\,,
\beq
\label{eq:MooseC-BK}
S_{\rm BT}^{~} ~= \dis\int\! d^5\xh~
\delta (x^5\!-\! L)
\f{\,-(\gh^{~}\!/g_{N+1}^{~})^2\,}{4}
\Fh^3_{\mu\nu}\Fh^{3\mu\nu} \,.
\eeq
This brane term is actually localized at $\,x^5=L^-$,\,
the left-neighborhood of \,$x^5=L$\, at which
the BC $\,\dif_5^{~}\Ah^{3}_\mu |_{x^5=L} \!=\,0\,$
is imposed, since our
deconstruction procedure of deriving this BC
from (\ref{eq:4D-BC2}) includes the 
boundary site $j=N+1$ plus its right-neighborhood.

\vspace*{3mm}
We then consider the Moose-D in Fig.\,6 
which is obtained from Fig.\,1 by folding
the $SU(2)$ lattice chain once at the middle
with renumbering \,$N\to 2N$\, and setting $\,M=0\,$.\,
The gauge symmetry breaking structure is
$\,SU(2)^{N+1}_L\otimes SU(2)_R^N \otimes U(1)
   \to U(1)_{\rm em}\,$,\, where 
$SU(2)_L^{N+1}$ part has inputs $\{g_j^{~},\,f_k^{~}\}$
and $SU(2)_R^{N}\otimes U(1)$ part has inputs 
$\{\gwt_j^{~},\,\fwt_k^{~}\}$.

\begin{figure}[t]
\label{fig:Fig6}
\begin{center}
\vspace*{-7mm}
\centerline{\psfig{file=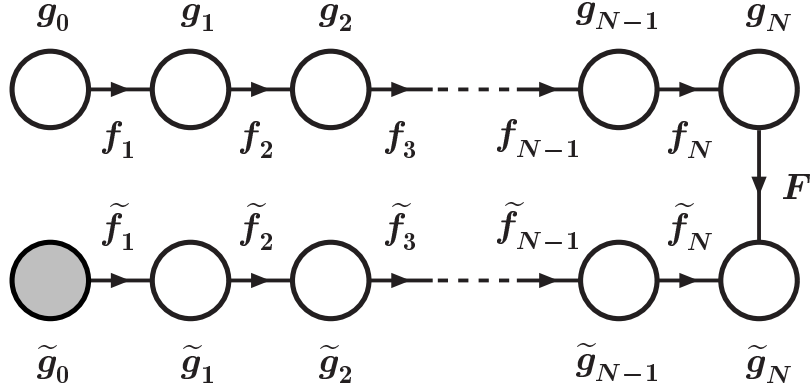,width=12cm}}
\setlength{\unitlength}{1mm}
\begin{picture}(0,0)
\end{picture}
\vspace*{-8mm}
\caption{Higgsless Moose-D:
Folding the $SU(2)^{N+1}$ moose chain once produces 
a variation of Higgsless deconstruction from Fig.\,1
with $\,M=0\,$ and by renumbering \,$N\to 2N$.\, 
The gauge symmetry breaking pattern is
$\,SU(2)^{N+1}_L\otimes SU(2)_R^N \otimes U(1)
   \to U(1)_{\rm em}\,$,\, 
whose continuum limit has a bulk gauge group
$\,[SU(2)_L\ot SU(2)_R]_{\rm 5D}\,$.}
\end{center}
\vspace*{-5mm}
\end{figure}

The continuum limit of Moose-D has the bulk gauge group
$\,[SU(2)_L\ot SU(2)_R]_{\rm 5D}\,$
(with 5D gauge couplings $\gh_L^{~}$ and $\gh_R^{~}$), 
plus a $U(1)_0$ brane term localized at \,$x^5=0$\,
[similar to (\ref{eq:MooseC-BK}) of Moose-C]. 
The other brane terms at \,$x^5=0$\, for $SU(2)_L$ 
[similar to (\ref{eq:MooseB-BK}) of Moose-C]
and at \,$x^5=L$\, for $SU(2)_D$ are also possible. 
From our moose formulation, 
it is clear that at \,$x^5=0$\,  
the BCs for $SU(2)_L$ and $SU(2)_R$ 
are essentially the same as those
of Moose-C at \,$x^5=0$\, and \,$x^5=L$,\, respectively,
i.e.,
\beq
\label{eq:5D-BC3-x50}
\left.\dif_5^{~}\Ah_L^{a\mu}\right|_{x^5=0} =0\,;
~~~~
\left.\Ah_R^{\pm\mu}\right|_{x^5=0} =0\,,
~~~~
\left.\dif_5^{~}\Ah_R^{3\mu}\right|_{x^5=0} =0\,.
\eeq
We then derive the new BCs at \,$x^5=L$\,.\,
Ordering the sites  
\,$j=0,1,\cdots,N,\linebreak
     \Nt,\widetilde{N\!\!+\!\!1},
     \cdots,\onet,\zerot$,\,
we can write down the charged gauge boson mass matrix 
$\,\MMW =\MMB_{[0,\zerot )}^2\,$ and the
neutral gauge boson mass matrix
$\,\MMZ =\MMB_{[0,\zerot ]}^2\,$.\,
Inspecting their eigenvalue equations we derive the
consistency relations for the eigenvectors 
\,$\XXB_n\in SU(2)_L$\, and \,$\XXBT_n\in SU(2)_R$,\, 
related to the boundary \,$x^5=L$\,,
\beq
\label{eq:4D-BC3}
\ba{rr}
\dis
F^2(g_N^{~}X_{N,n}-\gt_N^{~}\XT_{N,n}) +
f_{N+1}^2(g_{N+1}^{~}X_{N+1,n}-g_N^{~}X_{N,n}) & \,=~ 0 \,,
\\[2mm]
\dis
-F^2(g_N^{~}X_{N,n}-\gt_N^{~}\XT_{N,n}) +
\ft_{N+1}^2(\gt_{N+1}^{~}\XT_{N+1,n}-\gt_N^{~}\XT_{N,n}) & \,=~ 0 \,.
\ea
\eeq
The eigenvectors $\,(X_{jn},\,\XT_{jn})\,$ 
are related to the latticized 5D KK-wavefunctions 
$\,\dis \(\X_n(x^5_j),\,\XXT_n(x^5_j)\)\equiv (\X_{jn},\,\XXT_{jn})
=\sqrt{L}\(\f{g_j}{\gh_L}X_{jn},\,\f{\gt_j}{\gh_R}\XT_{jn}\)\,$.\, 
So we rewrite (\ref{eq:4D-BC3}) as conditions for 
\,$(\X_{jn},\,\XXT_{jn})$\,,
\beq
\label{eq:5D-BC3j}
\ba{rr}
\dis
F^2(\gh_L^{~}\X_{N,n}-\gh_R^{~}\XXT_{N,n}) +
\gh_L^{~}f_{N+1}^2(\X_{N+1,n}-\X_{N,n}) & \,=~ 0 \,,
\\[2mm]
\dis
-F^2(\gh_L^{~}\X_{N,n}-\gh_R^{~}\XXT_{N,n}) +
\gh_R^{~}\ft_{N+1}^2
(\XXT_{N+1,n}-\XXT_{N,n}) & \,=~ 0 \,,
\ea
\eeq
which are equivalent to
\beq
\label{eq:5D-BC3j2}
\ba{l}
\dis
\gh_L^{~}\X_{N,n}-\gh_R^{~}\XXT_{N,n}   ~=~
\gh_L^{~}(f_{N+1}/F)^2(\X_{N,n}-\X_{N+1,n})   \,,
\\[2mm]
\dis
\gh_R^{~} (\X_{N+1,n}-\X_{N,n}) + 
\zeta^2\,\gh_L^{~}(\XXT_{N+1,n}-\XXT_{N,n}) \,=~ 0 \,,
\ea
\eeq
where
$\,\zeta \equiv(\gh_R^{~}\ft_{N+1}^{~})/(\gh_L^{~}f_{N+1}^{~})\,$.\,
With the definitions $\,f_j^{~}= h_jf_L^{~}\,$ and
$\,\ft_j^{~}=h_jf_R^{~}\,$, we have the lattice spacing
$\,\aa \equiv (2/\gh_Lf_L)^2 = (2/\gh_Rf_R)^2$ [as shown above
(\ref{eq:SU2-5D-L})], which leads to a geometric equality
$\,\gh_L^{~}f_L^{~} = \gh_R^{~}f_R^{~}\,$.\, 
Hence the 5D geometry ensures $\,\zeta = 1\,$.\,
Taking continuum limit,
we derive the induced 5D BCs from (\ref{eq:5D-BC3j2}),
\beq
\label{eq:5D-BC3-x5L}
\left.\(\gh_L^{~}\Ah_L^{a\mu}
       -\gh_R^{~}\Ah_R^{a\mu}\)\right|_{x^5=L} =\,0\,,
~~~~~~
\left.\dif_5^{~}\!\(\gh_R^{~}\Ah_L^{a\mu}
                  + \gh_L^{~}\Ah_R^{a\mu}\)
                   \right|_{x^5=L} =\,0\,.
\eeq

We may also reconstruct a brane term at \,$x^5=L$\,.\,
This can be done by taking $\,f_N^2\gg f_{j<N}^2\,$
and $\,\ft_N^2\gg \ft_{j<N}^2\,$ together with
$\,F^2\to \infty\,$.\,
This will localize the diagonal group $SU(2)_N^D$
at \,$x^5=L$\,.\,
By definition, $\,g_j^{~}=\gh_L^{~}\ka_j/\sqrt{\aa}\,$
and $\,\gt_j^{~}=\gh_R^{~}\ka_j/\sqrt{\aa}\,$, which
contain the same $\ka_j$ because of the 5D geometry
in the continuum limit of Moose-D. Thus,
$\,\gh_L^{~}/g_N^{~}=\gh_R^{~}/\gt_N^{~}
  =\sqrt{\aa}/\ka_N^{~}\equiv\eta\,$.\,
So we can write the brane coupling of  $SU(2)_N^D$  as
$\,g_D^{~}=1/\sqrt{1/g_N^{2}+1/\gt_N^{2}}
=\eta^{-1}\gh_L^{~}\gh_R^{~}/\sqrt{\gh_L^2+\gh_R^2}\,$.\,
The brane gauge field is
$\,A^{a\mu}_{N,D}
=(\gt_N A^{a\mu}_N +g_N\At^{a\mu}_N)/\sqrt{g_N^2+\gt_N^2}\,$.\,
In the continuum limit, it becomes
$\,A^{a\mu}_{N,D}
=\eta (\gh_R^{~}\Ah^{a\mu}_L +
  \gh_L^{~}\Ah^{a\mu}_R )/\sqrt{\gh_L^2+\gh_R^2}\,$
at \,$x^5=L$\,,\,
and leads to the Yang-Mills term 
$\,\dis -\f{1}{4}(F_{N,D}^{a\mu\nu})^2 =
   -\f{\eta^2}{4(\gh_L^2+\gh_R^2)}
[\gh_R^{~}\Fh_L^{a\mu\nu}+\gh_L^{~}\Fh_R^{a\mu\nu}]^2 
$.~ 
Hence, we derive the brane localized kinetic term 
at \,$x^5=L$\,,
\beq
\label{eq:MooseD-BK}
S_{\rm BT} ~= \dis\int\! d^5\xh~ 
\delta (x^5\!-\! L)
\f{-\eta^2}{\,4(\gh_L^2+\gh_R^2)\,}
\[\gh_R^{~}\Fh_L^{a\mu\nu}+\gh_L^{~}\Fh_R^{a\mu\nu}\]^2 
\,.
\eeq
The possible brane terms at \,$x^5=0$\, can be similarly 
reconstructed as in our Moose-B and Moose-C.

\begin{figure}[h]
\label{fig:Fig7}
\begin{center}
\vspace*{-4mm}
\centerline{\psfig{file=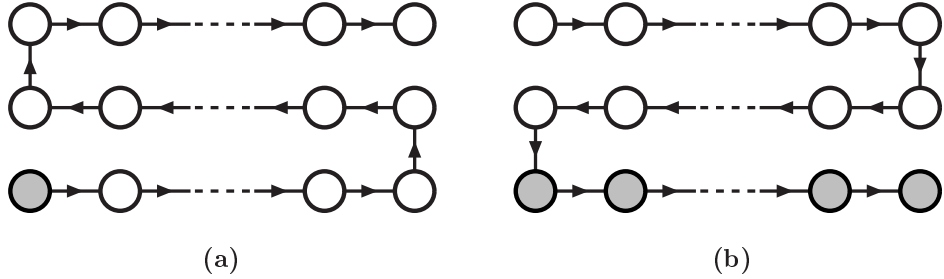,width=13cm}}
\setlength{\unitlength}{1mm}
\begin{picture}(0,0)
\end{picture}
\vspace*{-7mm}
\caption{Higgsless Moose-E with double-folding
from the general moose theory in Fig.\,1.
(a). Moose-E1 in the continuum limit has gauge group
$\,[SU(2)^3]_{\rm 5D}\ot U(1)_0\,$.\,
(b). Moose-E2 in the continuum limit has gauge group
$\,[SU(2)_L\ot SU(2)_R\ot U(1)_X]_{\rm 5D}\,$.
They are both broken to $U(1)_{\rm em}$.}
\end{center}
\vspace*{-4mm}
\end{figure}

A further extension of the Moose-D is to fold the
$SU(2)$-chain twice, called Moose-E1 in Fig.\,7(a).
This leads to a 5D continuum Higgsless theory with 
symmetry $\,[SU(2)^3]_{\rm 5D}\ot U(1)_0$\, whose BCs
can be similarly derived as in Moose-D. 
The gauge group breaks as 
$\,[SU(2)^3]_{\rm 5D}\ot U(1)_0 \to U(1)_{\rm em}$\,
under compactification.
Another variation is to add many $U(1)$'s to Moose-D
and form a separate $U(1)$-chain, 
or, from Fig.\,1 we can fold 
the $SU(2)$-chain once in its middle as well as 
folding the moose at
the intersection of the $SU(2)$ and $U(1)$ chains,
which we call Moose-E2 in Fig.\,7(b).
Its 5D continuum model has the
bulk gauge group 
$\,[SU(2)_L\ot SU(2)_R\ot U(1)_X]_{\rm 5D}$\,
which, for the case of warped geometry, gives precisely 
the 5D Higgsless model proposed by 
Cs\'{a}ki et al [\refcite{HLxx}].
Possible brane terms can be generated in
the continuum limit as we discussed before.
With the above method, 
we have also derived all the {\it induced} BCs in the
continuum limit of Moose-E2, 
and found that the BCs for 
$(\Ah^{a\mu}_L,\,\Ah^{a\mu}_R,\,\widehat{B}^{\mu})$ agree 
with the BCs of Cs\'{a}ki et al (based on the technique\cite{Heb} 
of adding extra boundary Higgs fields with 
large VEV).  
Systematic elaboration of these will be given 
elsewhere.\cite{HLmoosenew}

\section{$\!\!$Effective\,Unitarity\,of\,Higgsless\,Deconstruction\,Without~Geometry}

In this section we will demonstrate a conceptual point that 
the delayed unitarity violation (effective unitarity) in
the general moose theory can be realized without 
resorting to any known 5D geometry.
We observe that the delay of unitarity violation is
essentially a {\it collective effect} 
due to the participation in the EWSB  
from many gauge groups whose own symmetry breaking scales
$\{f_j^{~}\}$ are higher than the SM EWSB scale, 
\,$v=(\sqrt{2}G_F)^{-1/2}\simeq 246$\,GeV and whose gauge
couplings $\{g_j^{~}\}$ are larger than $g\in SU(2)_{\rm SM}$. 
Such a collective effect {\it does not} necessarily require
any exact 5D geometry, and can be realized in general 
non-geometric moose settings with relevant input parameters
$\,f_j^{~} > v\,$ and/or $\,g_j^{~} > g\,$.

\vspace*{-2mm}
\subsection{Formulation of the Effective Unitarity}

To analyze the unitarity it is enough to consider the
\,$g_{N+1}^{~}\to 0$\, limit, so the nontrivial part of Fig.\,1 
reduces to our Moose-B (Fig.\,3), corresponding to the
custodial symmetry limit. The other settings such as Moose-C,D,E 
also reduce to Moose-B at the zeroth order of \,$g_{N+1}^{~}$\,.\,
From Eq.\,(\ref{eq:L}) we first derive the relevant
unitary-gauge Lagrangian,
\beq
\label{eq:B-Lu}
\ba{ll}
{\cal L} & ~=~ \dis
\sum_{j=0}^{N} 
-\f{1}{4}{F}^a_{j\mu\nu}{F}_j^{a\mu\nu}
+\f{1}{2}{\AAB_\mu^a}^T\MMB_W^2\AAB^{a\mu}
\\[4mm]
& ~=~ \dis
\sum_{j=0}^{N} 
-\f{1}{4}\ov{W}^a_{j\mu\nu}\ov{W}_j^{a\mu\nu}
+\f{1}{2}{\WWB_\mu^a}^T{\MMB_W^2}^{\!\!\rm diag}\WWB^{a\mu}
+{\cal L}_G^{\rm int}    
\ea
\eeq
with the Yang-Mills interactions,
\beq
\label{eq:YM-int}
\ba{ll}
{\cal L}_G^{\rm int}   
& \dis ~=\, 
\sum_{j=0}^N \[
-\f{\,g_j^{~}}{2}C^{abc}
 \ov{F}^{a\mu\nu}_jA^b_{j\mu}A^c_{j\nu}
-\f{\,g_j^2}{4}C^{abc}C^{ade}
  A^{b\mu}_jA^{c\nu}_jA^d_{j\mu}A^e_{j\nu}
\]
\\[5mm]
& \dis ~=\,
-\f{\,G_3^{kmn}\,}{2}C^{abc}\ov{W}^{a\mu\nu}_kW^b_{m\mu}W^c_{n\nu}
-\f{\,G_4^{k\ell mn}\,}{4}C^{abc}C^{ade}
   W^{b\mu}_kW^{c\nu}_{\ell}W^d_{m\mu}W^e_{n\nu} \,,
\ea
\eeq
where 
$\,\ov{F}^{a\mu\nu}_j \equiv
   \dif^\mu A_j^{a\nu} -\dif^\nu A_j^{a\mu}\,$,\, 
$\,\ov{W}^{a\mu\nu}_j \equiv
   \dif^\mu W_j^{a\nu} -\dif^\nu W_j^{a\mu}\,$,\,
and
$\,\WWB^{a\mu}=(W^{a\mu}_0,\,\cdots,\,W^{a\mu}_N)^T\,$
denotes the mass-eigenbasis fields.
Also the sums over repeated indices 
\,$k,\ell,m,n=(0,1,\cdots, N)\,$ are implied.  
In Eq.\,(\ref{eq:YM-int})
the effective cubic and quartic gauge couplings are given by
\beq
\label{eq:g-3W-4W}
G_3^{kmn} \,=\,\sum_{j=0}^N g_j^{~}\,
             \RRB_{jk}^a\RRB_{jm}^b\RRB_{jn}^c\,,~~~~~~
G_4^{k\ell mn} \,=\,\sum_{j=0}^N g_j^2\,
             \RRB_{jk}^b\RRB_{j\ell}^d\RRB_{jm}^c\RRB_{jn}^e\,,
\eeq
with  \,$\RRB^{a} = \RRB_W$\,
determined from the diagonalization
$~\RRB_W^T\MMW\RRB_W\,= (\MMB_W^2\!)^{\rm diag}\,$.\,

The longitudinal gauge boson scattering 
$\,W^a_{mL}W^b_{nL}\to W^c_{kL}W^d_{\ell L}\,$
has the following structure,
\beq
\label{eq:WW-WW}
\ba{l}
\T^{ab,cd}_{mn,k\ell} ~ =
\T^{ab,cd}_{mn,k\ell [c]} +
\T^{ab,cd}_{mn,k\ell [s]} +
\T^{ab,cd}_{mn,k\ell [t]} +
\T^{ab,cd}_{mn,k\ell [u]}
\\[5mm]
~= \dis
G_4^{mnk\ell}\A^{abcd}_{[c]} + 
\sum_{j=0}^N \[
\f{G_3^{mnj}G_3^{k\ell j}}{\,s-M_{Wj}^2\,} 
\A^{abcd}_{[s]}  +  
\f{G_3^{mkj}G_3^{n\ell j}}{\,t-M_{Wj}^2\,} 
\A^{abcd}_{[t]}  +
\f{G_3^{m\ell j}G_3^{nkj}}{\,u-M_{Wj}^2\,} 
\A^{abcd}_{[u]} 
\],
\ea
\eeq
which is a sum of possible diagrams from the contact
interactions (denoted as $[c]$) 
and the gauge boson exchanges via
$(s,t,u)$-channels (denoted as $[s],[t],[u]$). 
Here the quantities \,$\A^{abcd}_{[c],[s],[t],[u]}$\,
contain an overall factor of relevant $SU(2)$ 
structure constants, and  
are functions of the energy $\sqrt{s}$ and 
scattering angle $\theta$, whose analytical
structures are identical to that of the (Higgsless) SM 
(with $g'\simeq 0$) up to a simple overall factor $g^2$.\, 
Taking the asymptotic energy \,$s\gg M_{Wj}^2$\, and
expanding the amplitude (\ref{eq:WW-WW}), 
we observe that all $O(E^4)$ terms exactly cancel 
as enforced by the ET (\ref{eq:ETm}),
while the nonzero $O(E^2)$ terms are {\it suppressed} by 
large \,$N\!+\!1$\,,\, leading to the {\it delayed unitarity
violation}\cite{CH}\,   in deconstruction theories.

To understand the suppressed $O(E^2)$ terms, it is very 
useful to examine the pure Goldstone interactions derived 
from (\ref{eq:L}), with general inputs for Moose-B, 
\beqa
\label{eq:L-pit-int}
{\mathcal L}_{\pit}^{\rm int} 
\,&=&\,\dis 
\f{~\CC_{k\l mn}~}{\,6v^2\,} \[
(\pit_k^a\partial_\mu\pit^a_{\l})
(\pit_m^b\partial^\mu\pit^b_{n})
- (\pit_k^a \pit_{\l}^a)
(\partial_\mu\pit_m^b \partial^\mu\pit_n^b)   \]
+O\!\(\pit_j^6\) ,
\eeqa
where \,$(k,\ell ,m,n)=0,1,2,\cdots,N$\, and
sum over repeated indices is implied.
The quartic coupling of the eaten Goldstone bosons is
\beqa
\label{eq:C}
\CC_{k\ell mn}^{~} ~&=&~ \dis
\sum_{j=1}^{N+1} \f{v^2}{\,f_j^2\,}
\RRBT_{jk}^a\RRBT_{j\ell}^a\RRBT_{jm}^b\RRBT_{jn}^b \leqq 1 \,,
\eeqa
with  \,$\RRBT^{a} = \RRBT_W$\,
determined from the diagonalization
$~\RRBT_W^T\MMWT\RRBT_W\,=\,
  (\MMBT_W^2\!)^{\rm diag}\,$.\,
For the special case of the Higgsless SM 
($N\!+\!1=M\!+\!1=1$),\, the coupling
$\,\CC_{0000}^{\rm SM} \equiv 1$\,,\, so there is
no suppression for the eaten Goldstone interactions. 
Computing the leading amplitude of the Goldstone scattering 
~$\pit_{n}^a \pit_{n}^b\to \pit_{n}^c\pit_{n}^d$\,
($n\geqq 0$),\,  we arrive at
\beq
\T[\pit_n^a\pit_n^b\to\pit_n^c\pit_n^d] ~=~
\dis\f{\,\CC_{nnnn}\,}{v^2}
\[s\,\d^{ab}\d^{cd} + t\,\d^{ac}\d^{bd} + u\,\d^{ad}\d^{bc}\],
\eeq
as compared to the unsuppressed result in the Higgsless SM,
\beq
\T[\pit_0^a\pit_0^b\to\pit_0^c\pit_0^d]_{\rm SM} ~=~
\dis\f{1}{\,v^2\,}
\[s\,\d^{ab}\d^{cd} + t\,\d^{ac}\d^{bd} + u\,\d^{ad}\d^{bc}\].
\eeq
Thus, we find that our general Moose-B has a {\it delayed}
unitarity violation scale for the Goldstone scattering 
~$\pit_{n}^a \pit_{n}^b\to \pit_{n}^c\pit_{n}^d$\, 
($n\geqq 0$),\,
relative to the \,$n=N=0$\, case of the Higgsless SM,
\beq
\label{eq:UB-pipi}
\dis
\Du \,\equiv\,\f{\Es}{\,\Essm\,} \,=\, 
\f{1}{~\sqrt{\CC_{nnnn}^{~}\,}~} > 1 \,,
~~~~~~{\rm with}~~~~~
\CC_{nnnn}^{~} \,=\,
\sum_{j=1}^{N+1} \f{v^2}{\,f_j^2\,}\,
\RRBT_{jn}^{a\,4}  \,.
\eeq
The original analysis
[\refcite{CH}] found that for a flat-geometry input  
\,$f_0^{~}=\cdots =f_N^{~}\equiv f$\, and
\,$g_0^{~}=\cdots =g_N^{~}\equiv\gt$\,,\,
the delay factor is
$\,\Du \simeq (f/v)\sqrt{N\!+\!1}\,$ 
for large $\,N\!+\!1$,\, 
where $f$ scales like\cite{CH}
$\,\sqrt{N\!+\!1}\,$ when holding $g$ and $1/R$\,.\,
This means\cite{CH} that for large $\,N\!+\!1$\,,\, 
the delay factor $\Du$ scales as \,$N\!+\!1$\,.

To analyze essential features of the delay factor
$\Du$, we first consider the simplest flat version of 
Moose-B (Fig.\,3) with massive zero-modes,
by setting the inputs
\,$f_0^{~}=\cdots =f_N^{~}=f_{N+1}^{~}\equiv f$\, and
\,$g_0^{~}=\cdots =g_N^{~}\equiv\gt$\,.\,
The Goldstone decay constants $f_j^{~}$ are connected to
the EWSB vacuum expectation value (VEV) $v$ via 
the low energy four-fermion interactions,\cite{HLmoose}
\beqa
\label{eq:v}
\[\sum_{j=p+1}^{N+1}\df{1}{\,f_j^2\,}\]^{-\f{1}{2}} 
&=&~ v ~=~ \(\sqrt{2}G_F\)^{-\f{1}{2}} 
\,\simeq~ 246\,{\rm GeV}\,,
\eeqa
where the site \,$j=p\in [0,N]$\, 
is the location of left-handed SM fermions.  
Thus, inputting $v$ makes one of $\{f_j^{~}\}$'s non-independent,
and for the flat (equal) $f_j^{~}$\,'s we deduce
$~f=v\sqrt{N+1-p~}~$.\,
The symmetry breaking pattern also imposes a relation
$\,g_0^{-2}+\cdots +g_N^{-2}=g^{-2}\,$  with 
$\,g\in SU(2)_W$,\, so for the flat (equal)  
$g_j^{~}$\,'s we have 
$\,g_j^{~}\equiv\gt =g\sqrt{N\!+\!1\,}\,$.\,
With these we exactly solve all the mass eigenvalues
and eigenvectors. In particular we find that
site-Goldstones \,$\Pi^a=(\pi_1^a,\cdots,\pi_{N+1}^a)^T$\,
are connected to the eaten Goldstones 
\,$\Pit^a=(\pit_0^a,\cdots,\pit_{N}^a)^T$\, by 
the orthogonal rotation $\,\Pi^a = \RRBT^a \Pit^a$\, with
\beq
\label{eq:MooseB-Rpi}
\RRBT_{jn}^a  ~=~ 
\sqrt{\f{2}{\,N+\f{3}{2}\,}}\sin\[(j+1)\alpha_n\],~~~~~
\( j,n=0,1,\cdots,N\) ,
\eeq  
where $\,\alpha_n \equiv \(n+\f{1}{2}\)\pi/\(N+\f{3}{2}\)\,$.\,
In the gauge sector we deduce the transformation from
the site-fields \,$\{A_j^{a\mu}\}$\, to the mass-eigenbasis fields
$\,\{W_n^{a\mu}\}\,$ via $\,\AAB^{a\mu}=\RRB^a\WWB^{a\mu}\,$
with
\beq
\label{eq:MooseB-Rw}
\RRB^a_{jn} = \dis
\sqrt{\f{2}{\,N+\f{3}{2}\,}}\cos\[\(j+\f{1}{2}\)\alpha_n\],~~~~~
\( j,n=0,1,\cdots,N \) .  
\eeq  
Substituting (\ref{eq:MooseB-Rpi}) and 
the relation $~f=v\sqrt{N+1-p\,}~$ into 
Eq.\,(\ref{eq:UB-pipi}), we compute the delay factor
for a flat Moose-B,
\beq
\Du \,=\, \dis
\f{f}{\,v\,}
\[\sum_{j=0}^{N}\RRBT_{jn}^{a\,4}\]^{-\f{1}{2}}
=~\[\f{2}{3}\(N+\f{3}{2}\)\(N+1-p\)\]^{\f{1}{2}} \,,
\eeq
which scales as \,$N/\sqrt{1.5}$\, for large
$\,N\!+\!1$\, and small \,$p =0-O(1)$\,.\,
For large $\,p\sim N\,$,\, the factor $\Du$ will be
reduced, scaling like $\,\sim\! \sqrt{N\,}\,$.\,

From the above, we see that a sizable delay factor 
$\,\Du > 1\,$
is essentially due to the {\it collective effect}
of the many gauge groups participating in the EWSB via 
{\bf (i)} 
the composition of the eaten Goldstone state 
$\pit^a_n=\sum_j\RRBT_{jn}^a\pi_j^a$
in terms of all $N\!+\!1$ site-Goldstones
$\{\pi^a_j\}$; and  
{\bf (ii)} 
many {\it comparable} symmetry breaking scales $\,f_j^{~}>v\,$ 
under the condition (\ref{eq:v}).
We observe that the realization of this collective effect
does not necessarily require the moose inputs 
\,$(g_j^{~},\,f_k^{~})$\, to obey any exact 5D geometry.
In fact, for the general parameters 
\,$(g_j^{~},\,f_k^{~})$\,,\, we can achieve a visible delay
factor $\,\Du > 1\,$ as long as $\,N\!+\!1\,$ is reasonably
large (implying that many site-Goldstones participate
in the collective symmetry breaking) and all relevant
$\,(g_j^{~},\,f_k^{~})\,$ are significantly larger than 
\,$(g,\,v)$\, of the SM.
With these two requirements satisfied, we find that, even if
the precise pattern of the inputs \,$(g_j^{~},\,f_k^{~})$\,
are very non-geometric and random-like, 
a nearly maximal delay factor can be reached for each
given \,$p$\,, which roughly scales like 
\beq
\label{eq:UB-max}
\dis
\Du ~\sim~
\Duu^{\max} ~\longrightarrow~ 
O(1)\[(N+1)(N+1-p)\]^{\f{1}{2}}  \,.
\eeq

Finally, we note that the above instructive estimate of the delay
factor $\Du$ is based on the leading Goldstone amplitude
at \,$O(E^2)$.\, 
The exact longitudinal gauge boson amplitude 
(\ref{eq:WW-WW}) may result in a stronger
unitarity limit $\Es$ and thus smaller delay factor
$\Du$, since the analysis of leading $E^2$-amplitude of
the Goldstone scattering has ignored all the subleading
terms of $O(E^0)$  and  $O(M_{Wn}^2/E^2)$ which
could be visible when $\,N\!+\!1\,$ is not too
large and the limit $\Es$ is only around a few TeV.
So, in the explicit analysis below,
we will directly compute the gauge boson amplitude
(\ref{eq:WW-WW}) for deriving the unitarity bound
\,$\Es$\,.

\subsection{Explicit Analysis of Effective Unitarity Without Geometry}

Next, we explicitly demonstrate how
the effective unitarity can be realized in various minimal
moose theories without geometry.  For the explicit analysis,
we will focus on the zero-mode amplitude
$\,\T[W^a_{0L}W^b_{0L}\to W^c_{0L}W^d_{0L}]\,$
which is also to be tested at the LHC.
The unitarity requires its $s$-wave amplitude to be
bounded from above,
$\,\left|a_{0000}^{0}[ab,cd]\right|< 1/2\,$,\,  
where the possible effect of final state identical particles
can be included\cite{DH}.\,    
We know that the naive Higgsless SM  
($N\!+\!1=M\!+\!1=1$) predicts the unitarity violation 
scale in the $W_{0L}W_{0L}$ scattering as, 
$\,E^\star_{\rm SM} \simeq \sqrt{8\pi}v\simeq 
   1.2\,$TeV\cite{SM-UB,CG,DH}.\,
This is significantly lower than the conventional cutoff
scale \,$\cut \simeq 4\pi v\simeq 3$TeV as estimated by
the consistency of chiral perturbation theory\cite{NDA}.\,
So we will identify and analyze the minimal moose (MM) models 
that can exhibit an effective unitarity at least 
up to a scale $E^\ast \gtrsim 3$\,TeV.

We perform the numerical analysis 
at the zeroth order of $U(1)$ coupling $g_{N+1}^{~}$. 
We ensure the light mass $M_W=M_{W0}\simeq 80$\,GeV 
and the lowest new gauge boson mass
$M_{W1}\gtrsim 800$\,GeV 
by the phenomenological consideration.  
We will consider one minimal moose $SU(2)^3\ot U(1)$ 
(called MM3, $N\!+\!1=3$ and $M=0$) with the 
left(right)-handed fermions 
coupled to the site $j=p=0$ ($j=q=3$) in Fig.\,1, 
and another minimal moose 
(called MM4, $N\!+\!1=4$ and $M=0$) with the left(right)-handed 
fermions coupled to the site $j=p=1$ ($j=q=4$).
We compute the unitarity limit $E^\star$ for
each model by scanning the parameter space in Fig.\,8, where the
$E^\star$ with ``non-geometric'' inputs 
(of random-like pattern) are shown by solid curves and 
are compared to the ``geometric'' inputs 
(with relevant $g_j^{~}$'s and/or 
$f_j^{~}$'s being equal) as depicted by dotted curves. 
We have chosen the scattering  
$\,W^+_{0L}W^-_{0L}\to W^+_{0L}W^-_{0L}\,$ in Fig.\,8
for illustration since other channels are found to
exhibit similar features.
For this process the $u$-channel term 
in (\ref{eq:WW-WW}) is absent. 
[The exact tree-level amplitude of this scattering in the
Higgsless SM ($N\!+\!1=M\!+\!1=1$) 
may be found in Eq.\,(3.17)
of the first paper in [\refcite{ET}] which does not pose
delayed unitarity violation.]
\begin{figure}[h]
\label{fig:8}
\vspace*{-6mm}
\centerline{\psfig{file=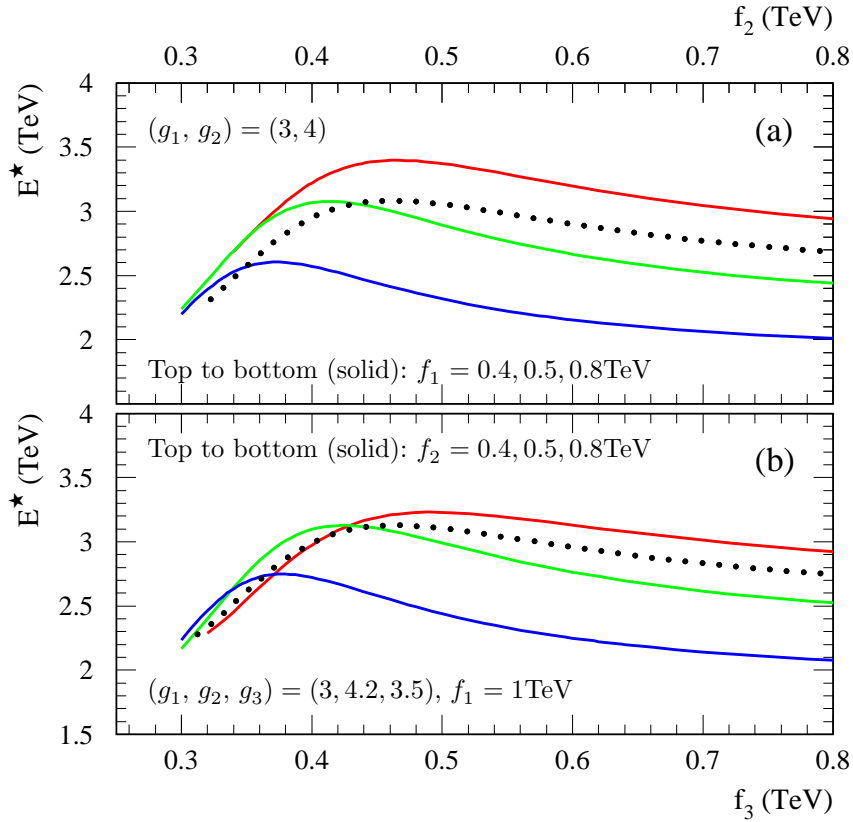,width=13cm}} 
\setlength{\unitlength}{1mm}
\begin{picture}(0,0)
\put(20,110){$(g_1^{~},\,g_2^{~})=(3,4)$}
\put(20,78){Top to bottom (solid): $f_1^{~}=0.4,0.5,0.8$TeV}
\put(20,35.4){$(g_1^{~},\,g_2^{~},\,g_3^{~})=(3,4.2,3.5)$, $f_1^{~}=1$TeV}
\put(20,67.5){Top to bottom (solid): $f_2^{~}=0.4,0.5,0.8$TeV}
\end{picture}
\vspace*{-18.5mm}
\caption{Unitarity limit $E^\star$ is shown 
for the moose $SU(2)^3\otimes U(1)$ (MM3) 
with $p=0$ in plot (a) as a function of $f_2^{~}$, 
and for the moose $SU(2)^4\otimes U(1)$ (MM4) with 
$p=1$ in plot (b) as a function of $f_3^{~}$.
The solid curves are shown for three sets of 
non-geometric inputs which appear random-like. 
As a comparison,
the dotted curves come from two sets of 
typical geometric inputs:
$g_1^{~}=g_2^{~}=4$, $f_1^{~}=\sqrt{3}v\simeq 0.43$\,TeV 
in plot (a); and 
$g_1^{~}=g_2^{~}=g_3^{~}=4$, 
$f_1^{~}=1$\,TeV, $f_2^{~}=\sqrt{3}v$ in plot (b).
}
\vspace*{-2mm}
\end{figure}     

Fig.\,8 shows that either solid or dotted curves 
are generally quite flat 
and there exists no sharp ``peak'' when we vary 
$\ffb$ in the plot (a) and $\ffc$ in the plot (b).
There is a sizable region on each curve to realize
a nearly maximal scale \,$\Es\sim \Es_{\max}$\,.\, 
The inputs for dotted curves mimic closely for
certain flat-geometry settings, but for realizing
\,$\Es\sim \Es_{\max}$\, 
they clearly do not exhibit any real
advantage over other non-geometric
inputs (shown as solid curves). 
We also find $\Es$ to be less sensitive to the variation 
of the pattern of gauge couplings.
To have a direct feeling, 
we explicitly list two sample inputs/outputs
for each model. For the MM3 model with $p=0$,
inputting \,$(g_1^{~},\,g_2^{~})=(3.1,\,4)$\, and
\,$(f_1^{~},\,f_2^{~})=(0.5,\,0.4)$\,TeV, we derive
the gauge boson mass-spectrum
$\,(M_{W0},\,M_{W1},\,M_{W2})\simeq (0.08,\,0.8,\,1.3)\,$TeV,
and the unitarity violation scale 
$\,\Es \simeq 3.1$\,TeV.  
For the MM4 model with $p=1$, inputting 
\,$(g_1^{~},\,g_2^{~},\,g_3^{~})=(3,\,4.2,\,3.5)$\, and
\,$(f_1^{~},\,f_2^{~},\,f_3^{~})=(1,\,0.5,\,0.38)$\,TeV, 
we derive the gauge boson mass-spectrum
$\,(M_{W0},\,M_{W1},\,M_{W2},\,M_{W3})\simeq 
   (0.08,\,0.81,\,1.3,\,1.8)\,$TeV,
and the unitarity violation scale 
$\,\Es \simeq 3.0$\,TeV; if we reduce the input values of
$\,(g_1^{~},\,g_2^{~})\,$ to
$\,(g_1^{~},\,g_2^{~})=(2.5,\,4)\,$ and keep other inputs
unchanged, we obtain a very similar mass-spectrum
$\,(M_{W0},\,M_{W1},\,M_{W2},\,M_{W3})\simeq 
   (0.08,\,0.8,\,1.2,\,1.6)\,$TeV and a unitarity limit
$\,\Es \simeq 3.1$\,TeV.

We have also analyzed other moose models with different 
inputs of \,$N\!+\!1$\, and/or \,$p$\, 
and found similar features, although the limit $\Es$
for each scattering process
generally increases for larger \,$N\!+\!1$\, and smaller
\,$p$\,,\, as expected. 
More systematic analysis including the effect of
delocalized fermions will be given elsewhere.\cite{HLmoosenew}

\section{Discussion and Conclusion}

In conclusion, 
using the powerful deconstruction approach has revealed that
4D gauge theories exhibit far richer dynamics than
naive expectation. 
With this conceptually clean tool,
we have formulated and classified 
the compactified 5D gauge symmetry
breaking (without/with gauge group rank reduction) 
via spontaneous symmetry breaking in 
general gauge-invariant 4D moose theories.
We demonstrate that
various consistent 5D boundary conditions (BCs) as well
as possible brane terms are automatically {\it induced}
from taking proper continuum limits of the general 
gauge-invariant moose theory.

We observe that the {\it effective unitarity} 
(delayed unitarity violation) 
is essentially a {\it collective effect} 
due to the participation in the EWSB  
from many gauge groups whose symmetry breaking scales
($f_j^{~}$)
are higher than the SM EWSB scale
$v=(\sqrt{2}G_F)^{-1/2}$ and whose gauge couplings
$g_j^{~}$ are larger than the SM $SU(2)_W$ coupling $g$\,.\,  
We find that the effective unitarity 
can be naturally realized in a wide class of 4D moose theories 
even with a few extra gauge bosons, 
without resorting to any known 5D geometry.
As shown in Sec.\,5, an important advantage of the 
advocated non-geometric moose theories is 
that they contain much larger parameter space than the
special geometric setting, so they open up much wider windows
for accommodating experimental constraints while retaining
effective unitarity.
More elaboration including the effect of delocalized fermions
will be given elsewhere\cite{HLmoosenew}.

The precision constraints on our general moose theory
(Fig.\,1) with the SM fermions localized
at arbitrary sites $j=p$ ($0\leqq p\leqq N$) and 
$j=q$ ($N+1\leqq q\leqq K+1$)
have been systematically analyzed\cite{HLmoose}.
In particular, by requiring a sizable gap between the 
light masses \,$M_{W,Z}$\, and the first heavy masses 
\,$M_{W',Z'} \,(\,\lesssim \sqrt{8\pi}\,v\,)$,\, 
a sum rule for the precision parameters
(corresponding to $\Sh$\,) can be derived,\cite{HLmoose}
\beq
\label{eq:Shat}
\!\!\Sh \,=\,\dis
\f{1}{4s_Z^2}\[\alpha S\!+\! 4c_Z^2\(\Delta\rho \!-\!\alpha T\)
\!+\!\f{\alpha\delta}{c_Z^2}\]
\geqq \sum_{r=p+1}^{N}\!\f{\,M_W^2\,}{M_r^2}
\,\gtrsim\, \f{M_W^2}{\,8\pi v^2\,}
\,\simeq\, 4\!\times\! 10^{-3},
\eeq
where $\,p\in [0,N)\,$ and  
$\,\{M_r^2\}$\, are eigenvalues of mass matrix
$\,\MMB_{(p,N+1)}^2$.\, 
Although the precision data seem not to favor (\ref{eq:Shat}), 
this phenomenological constraint may be evaded in a number 
of ways, {\it e.g.,} by smearing the
fermion location out of a single site in the general moose
theory,\cite{HLmoosenew}\, similar to the use of 
delocalized bulk fermions in certain 5D models\cite{HLyy}.

\vspace*{4mm}
\noindent
{\bf Acknowledgments}~\,
I am grateful to R. S. Chivukula, E. H. Simmons, M. Tanabashi and 
M. Kurachi for fruitful collaborations and helpful suggestions
to the manuscript.  
I also thank D. A. Dicus, A. Hebecker, R. Sundrum, J. Terning and  
U. Varadarajan for useful discussions. 
This work was supported by U.S. Department of Energy under
grant No.~DE-FG03-93ER40757.

\end{document}